\documentclass[12pt]{article}

\usepackage[dvipdfmx]{graphicx}

\usepackage{color,amsmath,slashed,booktabs,braket,hyperref,bm,authblk,graphicx,amssymb}
\usepackage[a4paper, total={16.5cm,22cm}]{geometry}
\RequirePackage[l2tabu, orthodox]{nag} 
\usepackage[all, warning]{onlyamsmath} 

\usepackage{newtxtext}
\usepackage[varg]{newtxmath} 
\usepackage[utf8x]{inputenc} 

\usepackage[sort&compress,numbers, merge]{natbib}

\hypersetup{
  colorlinks = true,
  linkcolor = blue,
  citecolor = blue,
}

\title{{\bfseries Cooling Theory Faced with Old Warm Neutron Stars: 
Role of Non-Equilibrium Processes with Proton and Neutron Gaps}}
\author[1]{Keisuke Yanagi}
\author[1]{Natsumi Nagata}
\author[1,2]{Koichi Hamaguchi}
\affil[1]{{\small \textit{Department of Physics, University of Tokyo, Tokyo 113--0033,Japan}}}
\affil[2]{{\small \textit{Kavli IPMU (WPI), UTIAS, The University of Tokyo, Kashiwa, Chiba 277--8583, Japan}}}
\date{}

\newcommand\Order{\mathop{\mathcal{O}}}

\newcommand\unit[1]{\,\mathrm{#1}}

\begin{document}
\hfill UT--19--06

\hfill IPMU19--0048

{\let\newpage\relax\maketitle}

\begin{abstract}

Recent observations have found several candidates for old warm neutron stars whose surface temperatures are above the prediction of the standard neutron star cooling scenario, and thus require some heating mechanism. Motivated by these observations, 
we study the non-equilibrium beta process in the minimal cooling scenario of neutron stars, which inevitably occurs in pulsars. This out-of-equilibrium process yields the late time heating in the core of a neutron star, called the rotochemical heating, and significantly changes the time evolution of the neutron star surface temperature. To perform a realistic analysis of this heating effect, we include the singlet proton and triplet neutron pairing gaps simultaneously in the calculation of the rate and emissivity of this process, where the dependence of these pairing gaps on the nucleon density is also taken into account. We then compare the predicted surface temperature of neutron stars with the latest observational data. We show the simultaneous inclusion of both proton and neutron gaps is advantageous for the explanation of the old warm neutron stars since it enhances the heating effect. It is then found that the observed surface temperatures of the old warm millisecond pulsars, J2124-3358 and J0437-4715, are explained for various choices of nucleon gap models. The same setup is compatible with the observed temperatures of ordinary pulsars including old warm ones, J0108-1431 and B0950+08, by choosing the initial rotational period of each neutron star accordingly. In particular, the upper limit on the surface temperature of J2144-3933 can be satisfied if its initial period is $\gtrsim 10\unit{ms}$. 
\end{abstract}


\thispagestyle{empty}
\clearpage

\section{Introduction}
\label{sec:introduction}

The cooling theory of a neutron star (NS) has been studied for over the years and tested against observations of the thermal emission from NSs~\cite{Yakovlev:1999sk, Yakovlev:2000jp, Yakovlev:2004iq}.
In particular, the minimal cooling paradigm~\cite{Page:2004fy, Gusakov:2004se, Page:2009fu} is successful in explaining these observations on the same footing.
It consists of the early time ($t\lesssim 10^5\unit{yr}$) neutrino emission from the core and late time ($t\gtrsim 10^5\unit{yr}$) photon emission from the surface. Fast neutrino emission processes such as the direct Urca process are not included---this assumption is valid unless NSs are very heavy~\cite{Lattimer:1991ib}. The minimal cooling takes account of the effect of the nucleon superfluidity\footnote{Throughout this paper, neutron superfluidity and proton superconductivity are collectively referred to as nucleon superfluidity.} as it plays important roles in NS cooling (see, e.g., Refs.~\cite{Page:2013hxa, Haskell:2017lkl, Sedrakian:2018ydt}). In the NS core, the spin-singlet pairing of protons and spin-triplet pairing of neutrons are formed once the internal temperature gets lower than the corresponding critical temperatures, giving rise to energy gaps in the spectra of elementary excitations of nucleons. Such an energy gap suppresses the slow neutrino emission~\cite{1994AstL...20...43L, 1995A&A...297..717Y, Gusakov:2002hh}.
In the meantime, the onset of nucleon pairings triggers the rapid neutrino emission from the Cooper pair breaking and formation (PBF) process~\cite{1976ApJ...205..541F, Voskresensky:1987hm, Senatorov:1987aa, Yakovlev:1998wr, Kaminker:1999ez, Leinson:2006gf}.
The formation of nucleon pairings also changes the contribution of nucleons to the heat capacity of the NS \cite{1994ARep...38..247L}.
In any case, the neutrino emission rate gets highly suppressed as the NS cools down, and eventually the surface photon emission dominates the neutrino emission, which then results in a rapid decrease in the NS surface temperature to $T_s^\infty \lesssim 10^4\unit{K}$ at $t\gtrsim 1~\unit{Myr}$. This is a generic consequence of the cooling theory, which can in principle be tested via the observation of the temperature of old NSs.

This minimal cooling theory is build upon the assumption that nucleons and leptons (electrons and muons) in the NS core are in chemical (or beta) equilibrium through the Urca reactions. This assumption, however, turns out to be invalid for (especially old) NSs.
As the rotation rate of a pulsar decreases, the centrifugal force decreases~\cite{Reisenegger:1994be}. 
This reduction makes the NS continuously contract, which perturbs the local number density of each particle species away from the equilibrium value. On the other hand, the timescale of the Urca reaction is typically much longer than that of the NS contraction (especially for old NSs), and thus the beta equilibrium cannot be maintained. This has a significant impact on the cooling of NSs, since the imbalance in the chemical potentials of nucleons and leptons, which quantifies the degree of the departure from beta equilibrium, is partially converted to the heat inside the NS~\cite{Reisenegger:1994be, 1992A&A...262..131H, 1993A&A...271..187G}. This heating mechanism due to the non-equilibrium Urca process is sometimes called the \textit{rotochemical heating}.
In the presence of the rotochemical heating, the NS surface temperature reaches $T_s^\infty \sim 10^5 \unit{K}$ at $t \gtrsim \unit{Gyr}$~\cite{Fernandez:2005cg}, which is in contrast to the prediction of the minimal cooling with the beta equilibrium. 

Intriguingly, several recent observations of old pulsars suggest that the late time heating operates in NSs. It is reported in Refs.~\cite{Kargaltsev:2003eb, Durant:2011je} that the (far) UV spectrum of the millisecond pulsar (MSP), J0437-4715, is consistent with the thermal emission with $T_s^\infty = (1.25-3.5) \times 10^5\unit{K}$.
The age of J0437-4715 is estimated independently from the pulsar spin-down age and the age of the companion white dwarf (WD), with both being compatible with the age of $t\simeq (6-7) \times 10^{9}\unit{yr}$. 
The surface temperature of $\sim 10^5\unit{K}$ for such an old NS cannot be explained by the cooling theory without heating.
Further evidences of old ``warm'' NSs are provided by the MSP J2124-3358~\cite{Rangelov:2016syg} and ordinary pulsars J0108-1431~\cite{Mignani:2008jr} and B0950+08~\cite{Pavlov:2017eeu}. On the other hand, the observation of the ordinary old pulsar J2144-3933 imposes an upper limit on its surface temperature, $T_s^\infty < 4.2 \times 10^4\unit{K}$ \cite{Guillot:2019ugf}, giving an evidence for the presence of an old ``cold'' NS. It is quite important to study if we can explain these observations by means of the minimal cooling with the non-equilibrium beta processes, i.e., with the rotochemical heating.\footnote{
  There are several other heating mechanisms to explain these warm NSs, including superfluid vortex creep~\cite{1984ApJ...276..325A, 1989ApJ...346..808S, 1991ApJ...381L..47V, 1993ApJ...408..186U, VanRiper:1994vp, Larson:1998it} and rotationally-induced deep crustal heating~\cite{Gusakov:2015kaa}.
  See also Ref.~\cite{Gonzalez:2010ta} for comparison of different internal heating mechanisms.
  }

To that end, it is necessary to include the non-equilibrium effect into the minimal cooling paradigm in a consistent manner. As mentioned above, the nucleon superfluidity plays a crucial role in the minimal cooling; this is also true for the rotochemical heating. The nucleon pairing gaps provide the threshold of the Urca reactions~\cite{Reisenegger:1996ir} because when the chemical imbalance is below this threshold, the Urca reactions are so suppressed that the heating is ineffective. In addition, the size of the threshold determines how much imbalance is to be generated, with
a larger imbalance enhancing the heating effect. The effect of nucleon superfluidity on the non-equilibrium beta processes has been considered in the literature \cite{Villain:2005ns, Petrovich:2009yh, Pi:2009eq, Gonzalez-Jimenez:2014iia}. In Refs.~\cite{Villain:2005ns, Pi:2009eq}, the rates and emissivities of the non-equilibrium beta processes were computed in the presence of either proton superconductivity or neutron superfluidity. In Ref.~\cite{Petrovich:2009yh}, the rotochemical heating was discussed for isotropic and uniform nucleon pairings. The anisotropic and density-dependent neutron pairing gaps were considered in Ref.~\cite{Gonzalez-Jimenez:2014iia}, but the effect of proton superconductivity was not included in the computation. In reality, however, \textit{both} the proton superconductivity and nucleon superfluidity are expected to occur in the NS core. The proton and neutron parings have the position- and temperature-dependence, and the neutron paring has also the angular-dependence. To perform a realistic simulation of the rotochemical heating, therefore, we need to include all of these effects simultaneously. Indeed, it is known in the calculation of the modified Urca processes that the simultaneous inclusion of all of these effects can considerably change the result \cite{Gusakov:2002hh}---this should also be the case for the non-equilibrium counterpart of the processes.

In this paper, we include both the proton singlet and neutron triplet pairings into the calculation of the NS thermal evolution in the presence of the non-equilibrium beta processes, with their density- and temperature-dependence fully taken into account. 
We then compare the results to the latest observational data of the NS surface temperature. 
We find that the heating with both nucleon pairings can be stronger than that only with neutron pairing~\cite{Gonzalez-Jimenez:2014iia}, which turns out to be advantageous in explaining the old warm MSPs. 

Meanwhile, it is found that the same setup can also account for the temperature data of the ordinary classical pulsars if the diversity of their initial spin periods is taken into account. 
We also discuss the compatibility of our scenario with the so-called X-ray dim Isolated Neutron Stars (XDINSs).

This paper is organized as follows.
We review the minimal cooling and rotochemical heating in Sec.~\ref{sec:therm-evol-neutr}.
In Sec.~\ref{sec:observations-old-hot}, we summarize the observational data of NS surface temperatures.
We then give our numerical analysis in Sec.~\ref{sec:results} and devote Sec.~\ref{sec:conclusions} to conclusions.
In appendix~\ref{sec:phase-space}, we summarize the formulas for phase space factors.
In Appendix~\ref{sec:magnetic-field}, we study the effect of varying magnetic field for the rotochemical heating.
\section{Thermal evolution of neutron stars}
\label{sec:therm-evol-neutr}
In this section, we review the theory of NS thermal evolution~\cite{Yakovlev:1999sk, Yakovlev:2000jp, Yakovlev:2004iq}.
Let us denote the local temperature in the core of a NS by $T(r)$, where $r$ denotes the distance from the NS center. This local temperature in general depends on the position, especially for a very young NS. It is however known~\cite{Yakovlev:1999sk, Yakovlev:2000jp, Yakovlev:2004iq} that the typical timescale of thermal relaxation in a NS is $\sim 10^{2-3}\unit{yr}$, and thus a NS with the age $t\gtrsim 10^4\unit{yr}$ can safely be regarded as isothermal. Since our main focus is on old NSs, in this work, we assume that NSs have already reached an isothermal state. In this case, the red-shifted internal temperature defined by $T^\infty \equiv T(r) e^{\Phi(r)}$ is constant throughout the NS core, with $e^{2\Phi(r)}=-g_{tt}(r)$ the time component of the metric.
The evolution of this red-shifted temperature is then governed by
\begin{align}
  \label{eq:time-evl}
  C\frac{dT^\infty}{dt} &= -L_\nu^\infty -L_\gamma^\infty + L_H^\infty,
\end{align}
where $L_\nu^\infty$ and $L_\gamma^\infty$ are the red-shifted luminosities of neutrino and photon emissions, respectively,
$C$ denotes the total heat capacity of the NS, and $L_H^\infty$ is the source of heating if exists.
In Sec.~\ref{sec:minimal-cooling}, we review the minimal cooling theory, assuming chemical equilibrium among nucleons and leptons. In this case, NSs just cool down due to the first two terms in the right-handed side of Eq.~\eqref{eq:time-evl}, with $L_H^\infty = 0$. We then discuss in Sec.~\ref{sec:rotochemical-heating} the internal heating effect caused by the non-equilibrium beta processes in rotating pulsars, which yields a non-zero $L_H^\infty$. 
In Sec.~\ref{sec:nucleon-gap-models}, we summarize the pairing gap models for nucleon superfluidity used in this work, which affect the evaluation of $L_\nu^\infty$ and $L_H^\infty$ significantly.

\subsection{Minimal cooling}
\label{sec:minimal-cooling}

The minimal cooling \cite{Page:2004fy, Gusakov:2004se, Page:2009fu} is a successful paradigm that can explain many NS surface temperatures. In this scenario, 
the energy loss of a NS for $t \lesssim 10^5\unit{yr}$ is caused by the neutrino emission from the core, whose dominant processes are the modified Urca and PBF: $L_\nu^\infty\simeq L_{\nu,M}^\infty + L_{\nu,\text{PBF}}^\infty$. The fast neutrino emission processes such as the direct Urca process are not included.  At later times, the surface photon emission becomes the dominant source for the NS cooling. In this subsection, we give a brief review on these processes. 

The modified Urca process consists of the reactions
\begin{align}
  &n + N_1 \to p + N_2 + \ell + \bar\nu_\ell \,,
  \label{eq:murca1} \\
  & p + N_2 + \ell \to n + N_1 + \nu_\ell\,,
    \label{eq:murca2}
\end{align}
where $N_1=N_2=n$ (neutron branch) or $N_1=N_2=p$ (proton branch) and $\ell = e, \mu$.
The emissivity, the energy loss rate per unit time and volume, of this process is given by
\begin{align}
  \label{eq:murca-emis}
  Q_{M,N\ell}
  &=
    \int \biggl[\prod_{j=1}^4 \frac{d^3p_j}{(2\pi)^3} \biggr] \frac{d^3p_\ell}{(2\pi)^3}\frac{d^3p_\nu}{(2\pi)^3}\,
    (2\pi)^4\delta^4(P_f - P_i) \cdot \epsilon_\nu \cdot\frac{1}{2} \sum_{\mathrm{spin}}|\mathcal M_{M,N\ell}|^2
    \notag\\
  &\times \left[f_1f_2(1-f_3)(1-f_4)(1-f_\ell) + (1-f_1)(1-f_2)f_3f_4f_\ell \right]\,,
\end{align}
where $j=1,2,3,4$ denote the nucleons $n,N_1,p, N_2$, respectively, $\delta^4(P_f - P_i)$ the energy-momentum conserving delta function, $1/2 \times\sum_{\mathrm{spin}}|\mathcal M_{M,N\ell}|^2$ the matrix element summed over all the particles' spins with the symmetry factor, and $f$'s the Fermi-Dirac distribution functions. In the NS core, nucleons and charged leptons are highly degenerate, and thus only those in a thermal shell near the Fermi surface can participate in the reaction, making the emitted neutrino energy $\epsilon_\nu$ of $\mathcal{O}(T)$. This is much smaller than the Fermi momenta of nucleons and charged leptons, and thus we can neglect the neutrino momentum in the momentum delta function as well as in the matrix element.
The matrix element is calculated in Ref.~\cite{1979ApJ...232..541F} based on the free one-pion exchange, with the angular dependence of the matrix element on the momenta of nucleons and leptons neglected. We also adopt this approximation in the following analysis.\footnote{
Recent development in this calculation suggests that there may be a significant change in the result if, e.g., the in-medium effects are included \cite{Blaschke:1995va, DehghanNiri:2016cqm, Schmitt:2017efp, Shternin:2018dcn}. We will study the impact of such effects on the calculation of the non-equilibrium beta processes on another occasion.}
We then obtain the total luminosity of the modified Urca process as
\begin{align}
  L_{\nu,M}^\infty &= \sum_{\ell=e,\mu}\sum_{N=n,p} \int dV \, Q_{M, N\ell} \,e^{2\Phi(r)}\,,
  \label{eq:non-eq-murca-luminosity}
\end{align}
where $dV=4\pi r^2 e^{\Lambda(r)}dr$ with $e^{2\Lambda(r)}=g_{rr}(r)$ being the radial component of the metric.

The effect of the nucleon superfluidity enters through the distribution functions. In the presence of nucleon superfluidity, the energy of a quasi-particle near the Fermi surface has the form
\begin{equation}
  \label{eq:spectrum}
  \epsilon_N(\bm p) \simeq \mu_N + \mathrm{sign}(p-p_{F,N})\sqrt{\Delta_N^2 + v_{F,N}^2(p-p_{F,N})^2}\,,
\end{equation}
where $\mu_N$ is the chemical potential, $p_{F,N}$ the Fermi momentum, and $v_{F,N}$ the Fermi velocity.
The energy gap $\Delta_N$ depends on the Fermi momentum $\bm p_{F,N}$ and temperature $T$: $\Delta_N= \Delta_N(\bm p_{F,N}, T)$. The gap develops a non-zero value below the critical temperature, which results in suppression in the emissivity roughly by a factor of $e^{-\Delta_N/T}$.
More detailed estimation of the suppression factor is given in Refs.~\cite{1995A&A...297..717Y, Gusakov:2002hh}, where the beta equilibrium (equilibrium between the reactions~\eqref{eq:murca1} and~\eqref{eq:murca2}) is assumed.
This assumption does not hold for spin-down pulsars, and the deviation from the beta equilibrium gives rise to heating in the NS core, as we see in Sec.~\ref{sec:rotochemical-heating}.

As mentioned above, the minimal cooling does not include the fast cooling processes such as the direct Urca process, which can occur only in the very dense region of a heavy NS~\cite{Lattimer:1991ib}.
For instance, the Akmal-Pandharipande-Ravenhall (APR) equation of state (EOS)~\cite{Akmal:1998cf} allows the direct Urca to operate for $M\gtrsim 1.97\,M_{\odot}$. We use this EOS in the following analysis. If such a fast cooling process occurs, it dominates the other neutrino emission processes and causes a rapid cooling of the NS. At present, the masses of most of the isolated NSs whose surface temperature is measured are unknown. Since our main focus is on NSs with relatively high surface temperatures, in this work we just assume that the direct Urca process does not occur in the NSs we consider---this is equivalent to the assumption that the NSs in question have masses $\lesssim 1.97\,M_{\odot}$. We however note in passing that an observation of NS surface temperature that is lower than our prediction may be explained just by including the direct Urca process. Of course, the validity of this assumption should be confirmed by measuring the NS masses in the future observations.

In addition to the suppression of the modified Urca process, the nucleon superfluidity triggers another source of neutrino emission called the PBF process:
\begin{align}
  \label{eq:pbf}
  \tilde{N}\tilde{N} \to [NN] + \nu_\ell + \bar\nu_\ell ~,
\end{align}
where $\tilde{N}$ represents a quasi-particle of the nucleon Fermi liquid and $[NN]$ indicates the nucleon pairing. This PBF process can occur only in the presence of nucleon superfluidity, and is active when the temperature is just below the critical temperature~\cite{Page:2009fu}; 
for lower temperatures, this process is again suppressed by a Boltzmann factor. It turns out that the PBF process tends to be dominant in the neutrino emission after the onset of nucleon superfluidity \cite{Gusakov:2002hh}. In particular, other neutrino emission processes such as bremsstrahlung are usually sub-dominant. In our analysis, we include only the dominant neutrino emission processes, i.e., the modified Urca and the PBF, and neglect the other processes such as bremsstrahlung, for simplicity.

After the PBF process becomes ineffective (typically for $t\gtrsim 10^5\unit{yr}$), the NS cools via the surface photon emission.
The photon emission obeys the Stefan-Boltzmann law: $L_\gamma=4\pi R^2\sigma_B T_s^4$, where $T_s$ is the surface temperature and $\sigma_B$ is the Stefan-Boltzmann constant. This surface temperature is different from the internal temperature as the envelope shields the surface from the interior. The relation between $T^\infty$ and $T_s$ depends on the constituent of the envelop, which is often parametrized by the ratio between the mass of the light elements in the envelop, $\Delta M$, and the NS mass, $M$: $\Delta M/M$.\footnote{The quantity $\eta \equiv g_{14}^2 \Delta M/ M$ is also used in the literature, where $g_{14}$ is the surface gravity in units of $10^{14}~\mathrm{cm}\cdot \mathrm{s}^{-2}$.}
We use the relation given in Ref.~\cite{Potekhin:1997mn} in the following analysis.\footnote{For more recent discussions on the envelope models, see Refs.~\cite{2016MNRAS.459.1569B, Wijngaarden:2019tht}. }

For the heat capacity $C$ in Eq.~\eqref{eq:time-evl}, we consider the contribution of neutrons, protons, electrons, and muons. The contribution of nucleons is modified after the nucleon superfluidity occurs, and is significantly suppressed when the temperature is much lower than its critical temperature~\cite{Yakovlev:1999sk}. In this case, the contribution of leptons dominates the total heat capacity of the NS.

All in all, by solving Eq.~\eqref{eq:time-evl} with $L_\nu^\infty$, $L_\gamma^\infty$, and $C$ given above, we obtain the temperature evolution of a NS. It is then found \cite{Page:2004fy, Gusakov:2004se, Page:2009fu} that the resultant temperature curves are compatible with the observed surface temperatures of NSs, except for several old warm NSs mentioned in the introduction. We will discuss in the subsequent sections that these exceptions can also be explained in the minimal setup, once the effect of the non-equilibrium beta processes is taken into account.

\subsection{Rotochemical Heating}
\label{sec:rotochemical-heating}
The rotochemical heating occurs in the very same setup as in the minimal cooling.
The only difference is whether the beta equilibrium of the Urca processes is assumed or not. In the case of an actual pulsar, its rotational rate keeps decreasing, which results in a continuous reduction of the centrifugal force. Consequently, NSs contract continuously and the equilibrium number densities of nucleons and charged leptons change at all times. The number densities of these particles in a NS follow the equilibrium values if the Urca reactions are fast enough. It however turns out that the typical timescale of these reactions is much longer than that of the NS contraction, especially for an old NS. As a result, the Urca reactions are no longer in beta equilibrium.
The departure from the beta equilibrium leads to an imbalance in the chemical potentials of nucleons and leptons, which keeps increasing over the time. The energy stored in the chemical imbalance is released partly via the neutrino emission and partly as a heat, generating a non-zero $L_H^\infty$ in Eq.~\eqref{eq:time-evl}. We emphasize that this heating effect due to the non-equilibrium beta reactions is an inevitable consequence for actual rotating NSs and thus needs to be included into the calculation of the temperature evolution of NSs. We review this heating mechanism in Sec.~\ref{sec:heating-rate}, and discuss the effect of superfluidity on the rotochemical heating in Sec.~\ref{sec:phase-space-factors}. 

\subsubsection{Heating rate}
\label{sec:heating-rate}

As in the minimal cooling, we consider a NS which is comprised of nucleons, electrons, and muons. We again assume that the fast neutrino emission processes do not operate in the NSs in question. As we discussed above, the Urca processes in spin-down pulsars are out of equilibrium, which leads to imbalance in the chemical potentials of nucleons and leptons. We denote this imbalance by 
\begin{equation}
    \eta_\ell \equiv \mu_n - \mu_p - \mu_\ell ~,
\end{equation}
for $\ell = e, \mu$. This parameter quantifies the departure from the beta equilibrium and is equal to the amount of the energy released in each reaction.

The neutrino emissivity in the out-of-equilibrium modified Urca process is again calculated by using Eq.~\eqref{eq:murca-emis}, with relaxing the assumption of the beta equilibrium. Following Refs.~\cite{Petrovich:2009yh, Gonzalez-Jimenez:2014iia}, we factorize the emissivity as
\begin{equation}
  \label{eq:non-eq-murca}
  Q_{M,N\ell} = Q_{M,N\ell}^{(0)} \, I^N_{M,\epsilon} ~,
\end{equation}
where $Q_{M,N\ell}^{(0)}$ denotes the emissivity computed for a non-superfluid NS in beta equilibrium. The dimensionless function $I^N_{M,\epsilon}$ represents the phase space integral, which incorporates the effect of the departure from beta equilibrium as well as that of nucleon superfluidity.

The departure from beta equilibrium generates heat due to the entropy production~\cite{Reisenegger:1994be, 1992A&A...262..131H, 1993A&A...271..187G}.
The heating luminosity is written as 
\begin{align}
  L_H^\infty &= \sum_{\ell=e,\mu}\sum_{N=n,p} \int dV\, \eta_{\ell} \cdot \Delta\Gamma_{M, N\ell} \, e^{2\Phi(r)}
               \label{eq:hating-rate}\,,
\end{align}
where $\Delta \Gamma_{M, N\ell}$ is the difference between the reaction rates of the processes~\eqref{eq:murca1} and~\eqref{eq:murca2}:
\begin{align}
  \label{eq:diff-rate}
  \Delta\Gamma_{M, N\ell}
  &=
    \int \biggl[\prod_{j=1}^4 \frac{d^3p_j}{(2\pi)^3} \biggr] \frac{d^3p_\ell}{(2\pi)^3}\frac{d^3p_\nu}{(2\pi)^3}\,
    (2\pi)^4\delta^4(P_f - P_i) \cdot \frac{1}{2} \sum_{\mathrm{spin}}|\mathcal M_{M,N\ell}|^2
    \notag\\
  &\times \left[f_1f_2(1-f_3)(1-f_4)(1-f_\ell) - (1-f_1)(1-f_2)f_3f_4f_\ell \right] ~.
\end{align}
We again factorize $\Delta \Gamma_{M,N\ell}$ by using the phase space factor $I^N_{M,\Gamma}$ as follows~\cite{Petrovich:2009yh, Gonzalez-Jimenez:2014iia}:
\begin{equation}
\label{eq:non-eq-murca_igamma}
    \Delta\Gamma_{M, N\ell} = \frac{Q_{M,N\ell}^{(0)}}{T(r)}\, I^{N}_{M,\Gamma}~.
\end{equation}
We study the phase space factors $I^N_{M,\epsilon}$ and $I^N_{M,\Gamma}$ in more detail in Sec.~\ref{sec:phase-space-factors}.

The time evolution of $T^\infty$ is still described by Eq.~\eqref{eq:time-evl}, now with a heating source $L_H^\infty$. The chemical imbalance parameters $\eta_\ell$ are also time-dependent. It is discussed in Ref.~\cite{Reisenegger:1996ir} that the diffusion timescale of the chemical imbalance is shorter than the time scale of its evolution, and thus we can safely assume that the red-shifted imbalance parameters $\eta_\ell^\infty \equiv \eta_\ell e^{\Phi(r)}$ are constant throughout the NS core. The evolution of $\eta_\ell^\infty$ is then determined by~\cite{Fernandez:2005cg}
\begin{align}
  \frac{d\eta_e^\infty}{dt}
  &=
    -\sum_{N=n,p}\int dV \,(Z_{npe}\Delta\Gamma_{M,Ne} + Z_{np}\Delta\Gamma_{M,N\mu})\, e^{\Phi(r)} +2W_{npe}\Omega\dot\Omega\,,
    \label{eq:roto-diff-eta-e}\\
   \frac{d\eta_\mu^\infty}{dt}
  &=
    -\sum_{N=n,p}\int dV \,(Z_{np}\Delta\Gamma_{M,Ne} + Z_{np\mu}\Delta\Gamma_{M,N\mu})\,e^{\Phi(r)} +2W_{np\mu}\Omega\dot\Omega\,,
    \label{eq:roto-diff-eta-mu}
\end{align}
where $\Omega(t)$ is the angular velocity of the pulsar.
$Z_{np}$, $Z_{np\ell}$ and $W_{np\ell}$ are the constants defined in Ref.~\cite{Fernandez:2005cg}, which depend on the stellar structure of the NS.
The first terms in the right-hand side of these equations describe the equilibration by the modified Urca process, while the second terms show the effect of spin-down, which makes the NS out of beta equilibrium. 

For the pulsar spin-down, we assume the power-law deceleration:
\begin{equation}
  \label{eq:pl-spin}
  \dot\Omega(t) = -k\Omega(t)^n\,,
\end{equation}
where $k$ and $n$ are positive constants.
Furthermore, we take $n=3$ in the following analysis; this braking index corresponds to the case where the loss of the rotational energy is caused solely by the magnetic dipole radiation.
In this case, we can solve Eq.~\eqref{eq:pl-spin} as
\begin{align}
  \label{eq:P-Pdot-dipole}
  \Omega(t) = \frac{2\pi}{\sqrt{P_0^2 + 2P\dot{P}\, t}}~,
\end{align}
where $P$ and $\dot P$ are the \textit{present values} of the pulsar period and its derivative, respectively,
and $P_0$ is the initial period $P(t=0)$.
Note that the combination $P \dot{P} = -  4\pi^2 \dot{\Omega}/\Omega^3 = 4\pi^2 k$ is time-independent for $n=3$ and related to the dipole magnetic filed; for instance, $B\sim 3.2\times 10^{19}\, (P\dot P/\mathrm{s})^{1/2}\unit{G}$ for a NS of radius $R=10\unit{km}$ and moment of inertia $I=10^{45}\unit{g}\unit{cm^2}$.
We also find from Eq.~\eqref{eq:P-Pdot-dipole} that if $P_0 \ll P$, then $t = P/(2\dot{P})$. This is called the spin-down age of a pulsar, which is widely used for the estimation of the pulsar age.

\subsubsection{Phase space factors}
\label{sec:phase-space-factors}

Now let us give more concrete expressions for the phase space factors $I^N_{M,\epsilon}$ and $I^N_{M,\Gamma}$ defined in Eqs.~\eqref{eq:non-eq-murca} and \eqref{eq:non-eq-murca_igamma}, respectively:\footnote{The prefactor is equal to the inverse of
\begin{equation}
    I_0 \equiv 2 \int_0^\infty dx_\nu \, x_\nu^3 \,
    \biggl[\prod_i^5 \int_{-\infty}^{\infty} dx_j\, f_j\biggr]\, \delta \biggl( \sum_{j=1}^5 x_j -x_\nu \biggr) 
    = \frac{11513 \pi^8}{60480} ~.
\end{equation}
We also note that $I^N_{M,\Gamma}$ in Eq.~\eqref{eq:i-integ-gamma} has the opposite sign to the corresponding phase space integrals given in Refs.~\cite{Petrovich:2009yh} and \cite{Gonzalez-Jimenez:2014iia}, while it is consistent with those in Ref.~\cite{Villain:2005ns}.
}
\begin{align}
  I^N_{M,\epsilon}
  &=
    \frac{60480}{11513\pi^8}
    \frac{1}{A_0^N}
    \int \prod_{j=1}^5\frac{d\Omega_j}{4\pi}
    \int_0^\infty dx_\nu \int_{-\infty}^{\infty}dx_ndx_pdx_{N_1}dx_{N_2} \,
    x_\nu^3 \cdot 
    f(z_n)f(z_p)f(z_{N_1})f(z_{N_2})\notag\\
  &\times\left[f(x_\nu - \xi_\ell -z_n-z_p-z_{N_1}-z_{N_2})
    + f(x_\nu + \xi_\ell -z_n-z_p-z_{N_1}-z_{N_2})\right]
    \delta^3\biggl(\sum_{j=1}^5\bm p_j\biggr)
    \,,
    \label{eq:i-integ-emis}
    \\
    I^N_{M,\Gamma}
  &=
    \frac{60480}{11513\pi^8}
    \frac{1}{A_0^N}
    \int \prod_{j=1}^5\frac{d\Omega_j}{4\pi}
     \int_0^\infty dx_\nu \int_{-\infty}^{\infty}dx_ndx_pdx_{N_1}dx_{N_2} \,
     x_\nu^2 \cdot 
    f(z_n)f(z_p)f(z_{N_1})f(z_{N_2})\notag\\
  &\times\left[f(x_\nu - \xi_\ell -z_n-z_p-z_{N_1}-z_{N_2})
    - f(x_\nu + \xi_\ell -z_n-z_p-z_{N_1}-z_{N_2})\right]
    \delta^3\biggl(\sum_{j=1}^5\bm p_j\biggr)
    \,,
    \label{eq:i-integ-gamma}
\end{align}
where we have defined 
\begin{equation}
    x_N \equiv \frac{\epsilon_N - \mu_N}{T}~, \quad 
    x_\nu \equiv \frac{\epsilon_\nu}{T}~, \quad 
    v_N \equiv \frac{\Delta_N(\bm p_{F,N}, T)}{T} ~, \quad 
    z_N \equiv \mathrm{sign}(x_N)\sqrt{x_N^2 + v_N^2} ~, \quad 
    \xi_\ell \equiv \frac{\eta_\ell}{T} ~,
\end{equation}
and $f(x) = 1/(e^x + 1)$. The constant factor $A_0^N$ is given by 
\begin{align}
    A_0^N  &\equiv \int \prod_{j=1}^5\frac{d\Omega_j}{4\pi} 
    \delta^3\biggl(\sum_{j=1}^5\bm p_j\biggr) 
    =
    \begin{cases}
    \frac{1}{8\pi p_{F,n}^3} & (N=n) \\
    \frac{(p_{F,\ell} + 3 p_{F,p} - p_{F,n})^2}{64\pi p_{F,n} p_{F,p}^3 p_{F,\ell}} 
    \Theta_p & (N=p)
    \end{cases}
    ~,
\end{align}
where $\Theta_p = 1$ for $p_{F,n}< 3 p_{F,p} + p_{F,\ell}$ and $0$ for otherwise, and
$j=1,2,\dots,5$ correspond to $n, p, \ell, N_1, N_2$, respectively. As can be seen from the definition in Eqs.~\eqref{eq:non-eq-murca} and \eqref{eq:non-eq-murca_igamma}, we have $I^N_{M,\epsilon}=1$ and $I^N_{M,\Gamma}=0$ in the limit of $v_n=v_p=\xi_\ell = 0$, i.e., for a non-superfluid NS in beta equilibrium.
The angular integral in the above equations is non-trivial in the presence of a neutron triplet pairing since its gap amplitude depends on the direction of $\bm p_{F,n}$ (see Sec.~\ref{sec:nucleon-gap-models}); otherwise it is just reduced to the factor $A_0^N$.

For superfluid matter, the phase space factors provide a threshold for the modified Urca process: $\eta_\ell \gtrsim 3\Delta_n + \Delta_p$ for the neutron branch and $\eta_\ell \gtrsim \Delta_n + 3\Delta_p$ for the proton branch~\cite{Reisenegger:1996ir}.
Hence, $\Delta_{\mathrm{th}} = \mathrm{min}\{3\Delta_n + \Delta_p, \Delta_n + 3\Delta_p\}$ is the threshold of the rotochemical heating---for $\eta_\ell \lesssim \Delta_{\mathrm{th}}$, heating does not occur because the modified Urca reaction is suppressed ($\Delta\Gamma_{M,N\ell}\simeq0$). For a very young NS, the Urca reaction is fast enough so that the chemical equilibrium is maintained, i.e., $\eta_\ell = 0$. Later, the NS departs from beta equilibrium due to the spin-down and $\eta_\ell$ monotonically increases until it exceeds $\Delta_{\mathrm{th}}$, after which the accumulated $\eta_\ell$ is converted to heat.\footnote{We however note that if $\Delta_{\mathrm{th}}$ is large and/or the increase rate of $\eta_\ell$ is small, $\eta_\ell$ may never exceed the threshold and thus rotochemical heating is always ineffective. We also see such situations in Sec.~\ref{sec:ordinary-pulsars}.}
Therefore, the rotochemical heating is efficient usually at late times, when $T \ll \Delta_N$ and $T \ll \eta_\ell$. In such a situation, we can safely exploit the zero temperature approximation in the calculation of the phase space factors \cite{Petrovich:2009yh}.

A numerical calculation in Ref.~\cite{Fernandez:2005cg} shows that the late-time heating indeed occurs in a non-superfluid NS. In this case, we have analytical expressions for the phase space factors~\cite{Reisenegger:1994be}. Such analytical expressions for the neutron (proton) branch can also be obtained for the case where only protons (neutrons) form a constant paring gap in the limit of $T\to 0$~\cite{Petrovich:2009yh}. As for the numerical evaluation of the phase space factors, Refs.~\cite{Villain:2005ns, Pi:2009eq} give the results for the case in which either proton or neutron has a non-zero gap. Reference~\cite{Petrovich:2009yh} also performs the numerical computation of the phase space factors in the presence of the neutron and proton singlet uniform pairings.
In Ref.~\cite{Gonzalez-Jimenez:2014iia}, neutron triplet pairings whose gap has density and temperature dependence are considered, but the effect of proton superfluidity is neglected.
In this work, we include the effect of both the singlet proton and triplet neutron pairing gaps with taking account of their density and temperature dependence. For the calculation of the phase space factors, we use the zero temperature approximation as in Ref.~\cite{Petrovich:2009yh}; see Appendix~\ref{sec:phase-space} for more details.

\subsection{The nucleon gap models}
\label{sec:nucleon-gap-models}


The strength of neutrino emission and the rotochemical heating depends on the size of nucleon gaps through the phase space factors. 
In the NS core, singlet pairing of protons and triplet pairing of neutrons are formed once the internal temperature has fallen below their critical temperatures. For proton singlet pairing, the gap is independent of the orientation of the proton momentum $\bm p_{F,p}$.
On the other hand, neutron triplet gap is dependent on the angle $\theta$ between the neutron momentum $\bm p_{F,n}$ and the quantization axis. This angular dependence is different among states which have different projections $m_J$ of the total angular momentum on the quantization axis. Currently, it is unclear which of the states is actually formed in the NS core. In this work, we just consider the $m_J = 0$ state, which has the angular dependence of $\Delta_N(\bm p_{F,N}, T) \propto \sqrt{1+3\cos^2\theta}$. The application of our analysis to the case of the $|m_J| = 2$ state is straightforward.\footnote{See Ref.~\cite{Gonzalez-Jimenez:2014iia} for a comparison of the rotochemical heating in the cases of $m_J=0$ and $|m_J|=2$.} 

\begin{figure}
  \centering
  \begin{minipage}{0.5\linewidth}
    \includegraphics[width=1.0\linewidth]{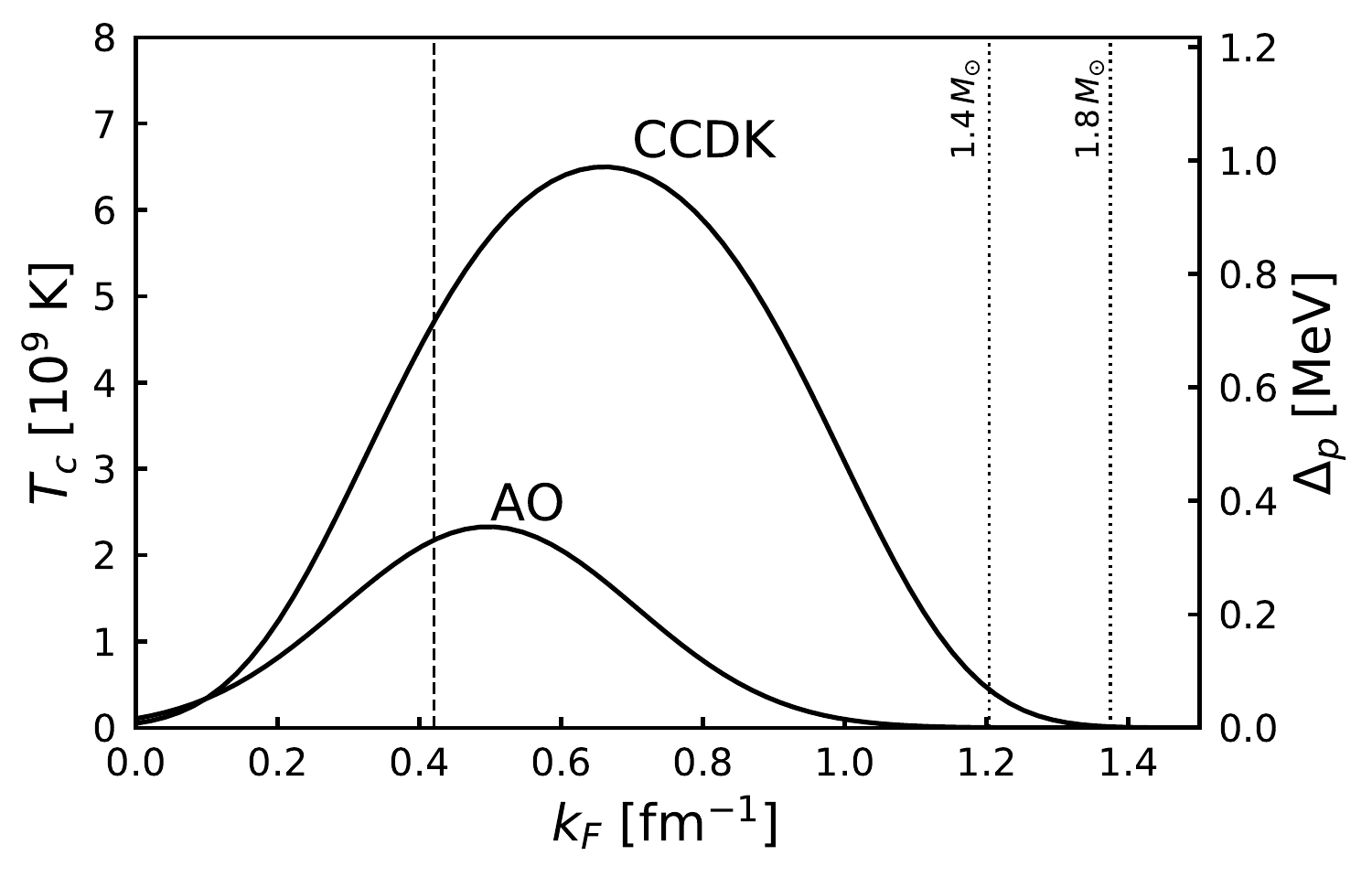}
  \end{minipage}%
  \begin{minipage}{0.5\linewidth}
    \includegraphics[width=1.0\linewidth]{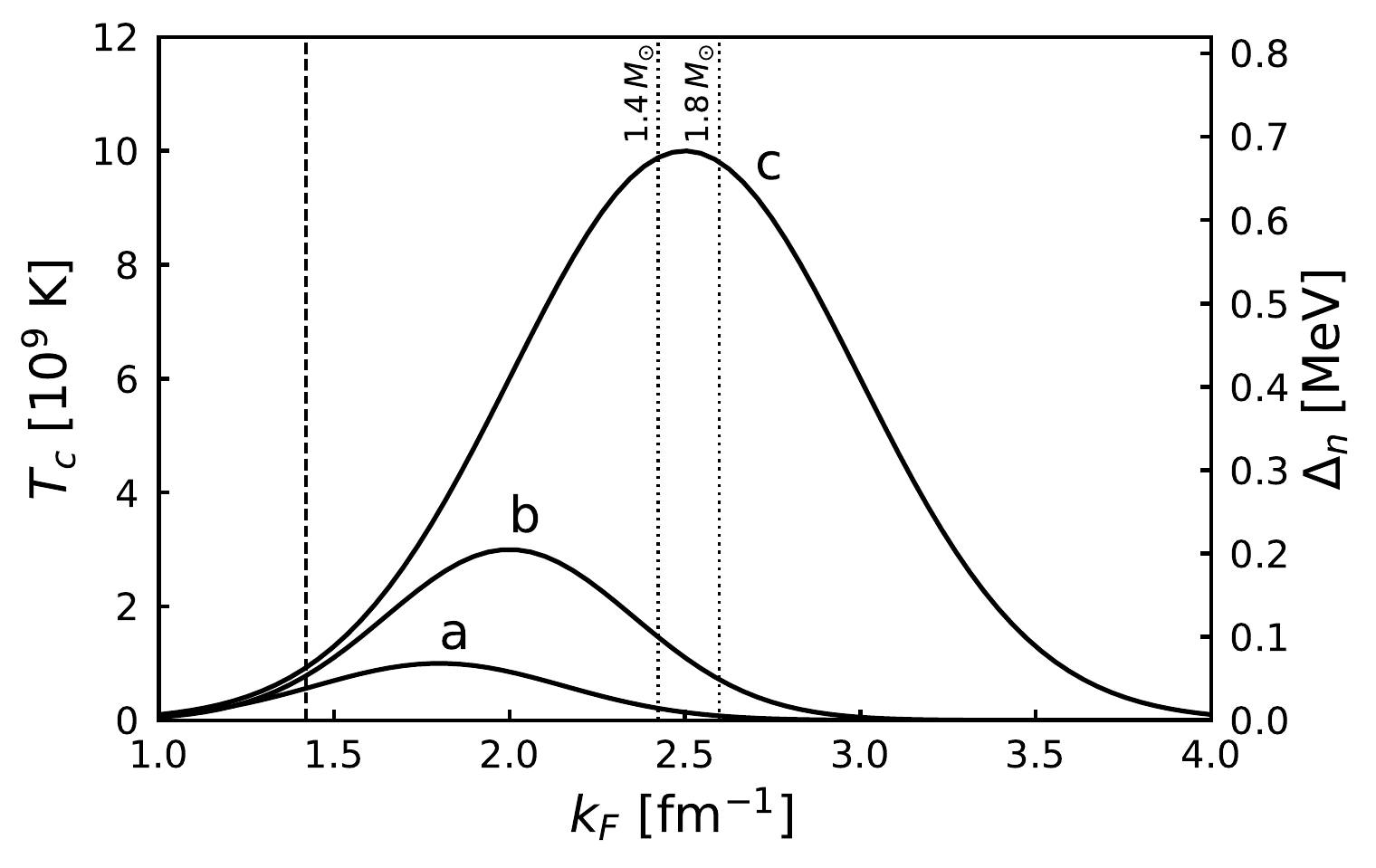}
  \end{minipage}
  \caption{Nucleon gaps used in our analysis; Left: proton $^1\mathrm{S}_0$ gap models; Right: neutron $^3\mathrm{P}_2~(|m_J|=0)$ gap models.
    The critical temperature (left axis), as well as the corresponding zero temperature gap (right axis), of each model is plotted as a function of the Fermi wave number $k_F$.
    The right vertical axis in the right panel corresponds to the gap amplitude for $\cos\theta = 0$.
    The dashed lines show the boundary between the crust and core.
  The dotted lines correspond to the NS center density for $M=1.4\,M_\odot$ and $1.8\,M_\odot$.}
  \label{fig:gap}
\end{figure}

Calculation of gap profile suffers from uncertainty due to the nucleon interactions and medium effects. There are many theoretical calculations/models of the zero temperature gap, $\Delta_N(\bm p_{F,N}, T=0)$, which is proportional to the critical temperature.
Given the zero temperature gap, we can obtain the gap amplitude for $T\neq 0$ by using the fitting formula in Ref.~\cite{Yakovlev:1999sk}.
For the proton gap, several theoretical models are shown in Ref.~\cite{Page:2013hxa}, among which we use the AO model~\cite{Amundsen:1984qq} as a small gap and the CCDK model~\cite{Chen:1993bam} as a large gap.
The calculation of neutron triplet gap is more uncertain, so we just use the phenomenological models named ``a'', ``b'' and ``c'' in Ref.~\cite{Page:2004fy}, for which the critical temperature is given by a Gaussian-shape function with respect to the Fermi wave number $k_F \equiv p_{F,N}/\hbar$:
\begin{equation}
    T_c (k_F) = T_0 \exp \biggl[- \frac{(k_F - k_0)^2}{\Delta k^2}\biggr] ~,
\end{equation}
with $T_0=1.0\times 10^9\unit{K}$, $k_0 = 1.8\unit{fm}^{-1}$, and $\Delta k = 0.5\unit{fm}^{-1}$ for ``a'', $T_0=3.0\times 10^9\unit{K}$, $k_0 = 2.0\unit{fm}^{-1}$, and $\Delta k = 0.5\unit{fm}^{-1}$ for ``b'', and $T_0=1.0\times 10^{10}\unit{K}$, $k_0 = 2.5\unit{fm}^{-1}$, and $\Delta k = 0.7\unit{fm}^{-1}$ for ``c''. In Fig.~\ref{fig:gap}, we show the critical temperature/zero temperature gap of the models used in our analysis as a function of the Fermi wave number $k_F$. As for the neutron triplet pairing gap, we show the amplitude for $\cos \theta = 0$.
The dashed line indicates the core-crust boundary and the dotted lines correspond to the NS center for $M=1.4\,M_\odot$ and $1.8\,M_\odot$. As can be seen from the plots, even at low temperatures,
protons can be in the normal fluid state near the NS center depending on the NS mass and gap models, while neutrons are in the superfluid state in the entire core for any gap model.

\section{Observations of neutron star temperatures}
\label{sec:observations-old-hot}

\begin{table}
  \centering
  \caption{Observational data used in this work. $t_{\text{sd}}$, $t_{\text{kin}}$, 
  $T_s^\infty$, and $P$ denote the spin-down age, kinematic age, 
  effective surface temperature, and period of neutron stars, respectively. 
  The sixth column shows the atmosphere model used in the estimation of 
  the surface temperature, where H, BB, C, and PL indicate hydrogen,  
  blackbody, carbon, and power-law, respectively, while M represents a magnetized NS hydrogen atmosphere model, such as NSA \cite{1995ASIC..450...71P} and NSMAX \cite{Ho:2008bq}. Data are taken from ATNF Pulsar
  Catalogue \cite{Manchester:2004bp, atnf} unless other references are shown explicitly.}
  \vspace{3mm}
  {
  \begin{tabular}{lccccc}\hline\hline
    Name & $\log_{10}t_{\mathrm{sd}}$& $\log_{10}t_{\mathrm{kin}}$ & $\log_{10}T_s^\infty$ &$P$   &Atmos. \\ 
    &[yr]&[yr] &[K] &[s] &model \\\hline
    PSR J2124-3358  & $10.0^{+0.2}_{-0.1}$ \cite{Rangelov:2016syg}&&$4.7-5.3$ \cite{Rangelov:2016syg} &$4.9\times 10^{-3}$& BB+PL \\
    PSR J0437-4715  & $9.83^{+0.01}_{-0.02}$ \cite{Durant:2011je}&& $5.1-5.5$ \cite{Durant:2011je}&$5.8\times 10^{-3}$& BB \\
 \hline
    PSR J2144-3933  &$8.5$&&$<4.6$ \cite{Guillot:2019ugf}&$8.5$& BB\\
    PSR J0108-1431  &$8.3$ \cite{2009ApJ...701.1243D}&&$5.0-5.7$ \cite{Mignani:2008jr}&$0.81$& BB  \\
    PSR B0950+08  &$7.2$&&$5.0-5.5$ \cite{Pavlov:2017eeu}&$0.25$& BB+PL  \\
    RX J2143.0+0654  &$6.6$&&$6.082 (2)$ \cite{Kaplan:2009au} &$9.4$&BB\\
    RX J0806.4-4123 &$6.5$&&$6.005(5)$ \cite{Kaplan:2009ce} &$11.4$& BB \\
    PSR B1929+10 &$6.5$&&$<5.7$ \cite{Becker:2005dk} &$0.23$& BB+PL \\
    RX J0420.0-5022 &$6.3$&&$5.742(3)$ \cite{Kaplan:2011xd}&$3.5$& BB\\
    PSR J2043+2740  &$6.1$&&$5.64(8)$ \cite{Beloin:2016zop}&$0.096$&H\\
    RX J1605.3+3249 &$4.5$ \cite{Pires:2014qza}&$5.6-6.6$ \cite{Tetzlaff:2012rz}&$5.88(1)$ \cite{Pires:2019qsk}&$3.4$ \cite{Pires:2014qza}& BB \\
    RX J0720.4-3125 &$6.3$& $5.8-6.0$ \cite{Tetzlaff:2011kh} &$5.5-5.7$ \cite{Kaplan:2003hj}&$8.4$&BB\\
    RX J1308.6+2127 &$6.2$& $5.5-6.2$~\cite{Motch:2009nq}&$6.07(1)$ \cite{Schwope:2006ra}&$10.3$& BB \\
    PSR B1055-52 &$5.7$&&$5.88 (8)$ \cite{Beloin:2016zop}&$0.20$&BB\\
    PSR J0357+3205 &$5.7$&&$5.62^{+0.09}_{-0.08}$ \cite{Kirichenko:2014ona}&$0.44$&M+PL\\
    RX J1856.5-3754 &$6.6$&$5.66(5)$ \cite{Tetzlaff:2011kh} &$5.6-5.7$ \cite{Sartore:2012fk}&$7.1$&BB \\
    PSR J1741-2054 &$5.6$&&$5.85^{+0.03}_{-0.02}$ \cite{Auchettl:2015wca}&$0.41$&BB+PL\\
    PSR J0633+1748 &$5.5$&&$5.71(1)$ \cite{Mori:2014gaa}&$0.24$&BB+PL\\
    PSR J1740+1000 &$5.1$&&$6.04(1)$ \cite{2012Sci...337..946K}&$0.15$&BB\\
    PSR B0656+14  &$5.0$&&$5.81(1)$ \cite{DeLuca:2004ck}&$0.38$&BB+PL\\
    PSR B2334+61 &$4.6$&&$5.5-5.9$ \cite{McGowan:2005kt}&$0.50$&M\\
    PSR J0538+2817 &$5.8$& $4.3-4.8$ \cite{Ng:2006vh}&  $6.02(2)$ \cite{Ng:2006vh}&$0.14$&H\\
    XMMU J1732-344 &&$4.0-4.6$ \cite{Tian:2008tr, Klochkov:2014ola} &$6.25(1)$ \cite{Klochkov:2014ola}&&C\\
    PSR B1706-44   &$4.2$ & & $5.8^{+0.13}_{-0.13}$ \cite{McGowan:2003sy}&$0.10$&M+PL\\
    \hline\hline
  \end{tabular}
  }
  \label{tab:psr-temp}
\end{table}

Before going to our numerical study, we summarize the current status of the NS temperature observations.
We focus on isolated NSs with the age $t>10^4\unit{yr}$, for which we can safely assume that the thermal and diffusion relaxation in the NS core has already been completed. 
In Tab.~\ref{tab:psr-temp}, we list the NSs whose surface temperature is measured, together with two NSs (PSR J2144-3933 and PSR B1929+10) for which only the upper bound on the surface temperature is obtained.
For most of the NSs in the table, only the spin-down age $t_{\mathrm{sd}}=P/(2\dot P)$ can be used for the estimation of their age, while in some cases, the kinematic age $t_{\mathrm{kin}}$, which is derived from the motion of the supernova remnant, is also available. 
In this work, we use the kinematic age if available, and the spin-down age otherwise.%
\footnote{We note that once we fix the initial period and the value of $P \dot{P}$ (hence the dipole magnetic field), the NS age is unambiguiously determined from the observed value of the NS period through Eq.~\eqref{eq:P-Pdot-dipole}. In particular, the spin-down age should agree to the real pulsar age without uncertainty if $P_0 \ll P$.  } The values of $t_{\mathrm{sd}}$ and $P$ are taken from ATNF Pulsar Catalogue \cite{Manchester:2004bp, atnf} unless other references are shown explicitly.

The surface temperature is estimated from the fit of the observed spectrum with atmosphere models.
Generically speaking, the hydrogen/carbon atmosphere model fits the thermal emission of young pulsars ($t \lesssim 10^5\unit{yr}$) well, while the heavy element model or blackbody offers a good fit for a middle-aged or old pulsar. In addition, if a NS has a relatively large magnetic field, a magnetized NS hydrogen atmosphere model such as NSA \cite{1995ASIC..450...71P} and NSMAX \cite{Ho:2008bq} may improve its spectrum fit. In the sixth column in Tab.~\ref{tab:psr-temp}, we show the atmosphere model used in the evaluation of the surface temperature of each NS; H, BB, C, PL, and M represent the hydrogen, blackbody, carbon, power-law, and magnetized NS hydrogen atmosphere models, respectively. For some NSs in the table, there are several atmosphere models that can fit the observed spectrum. In such cases, we choose the model which is considered to give the best fit and/or yields a NS radius of a plausible size ($\sim 10$~km). Moreover, there are some cases where the use of two or more blackbody components improves the fit due to the presence of hot spots, though we do not show this explicitly in the table. In these cases, we use the temperature of the component with a NS radius that is consistent with the typical NS radius. 
Note, however, that the fits with atmosphere models are often performed with the mass, radius, or distance of the NS being fixed, though these quantities are not precisely known for most of the NSs---if this is the case, additional systematic uncertainty might be present. For more details, see the references cited in the fourth column of the table.

The first two pulsars in Tab.~\ref{tab:psr-temp}, PSR J2124-3358 and J0437-4715, are classified into the MSPs.
They have small $P$ and $\dot P$, and hence a small dipole magnetic field, compared to the ordinary (classical) pulsars.
PSR J0437-4715 is the closest millisecond pulsar at present.
Its rotational period, mass, and distance 
are estimated to be $5.76$~ms, 
$1.44 \pm 0.07~M_{\odot}$, and $d= 156.79 \pm 0.25$~pc, 
respectively \cite{Reardon:2015kba}.
This pulsar is in a binary system accompanied with a white 
dwarf. The spin-down age of PSR J0437-4715 is estimated in 
Ref.~\cite{Durant:2011je} to be $t_{\text{sd}} = (6.7\pm 0.2)
\times 10^9$~years with the Shklovskii correction 
\cite{1970SvA....13..562S} included. This is in a 
good agreement with
the estimated age of the white dwarf, $t_{\text{WD}} = 
(6.0\pm 0.5) \times 10^9$~years. 
In Ref.~\cite{Durant:2011je}, it is found 
that a fit with the Rayleigh-Jeans law in the far UV range 
is consistent with a blackbody emission from the whole
NS surface with a temperature of 
$T_s^\infty = (1.25-3.5) \times 10^5$~K. As argued in 
Ref.~\cite{Durant:2011je}, it is unlikely for the observed
surface temperature to be due to heat flow coming from
the magnetosphere regions. 
Since the minimal cooling theory predicts $T_s^\infty \ll 10^3\unit{K}$ for $t\sim 10^{10}\unit{yr}$, the observed surface temperature requires late time heating.
PSR J2124-3358 is an isolated millisecond pulsar with a period of $4.93$~ms \cite{Reardon:2015kba}. Its spin-down age, 
after corrected by the Shklovskii effect, is 
$11^{+6}_{-3}\times 10^9$~years \cite{Rangelov:2016syg} for 
the distance $d = 410^{+90}_{-70}$~pc. Its surface temperature is obtained with a blackbody plus power-law fit to be $(0.5-2.1) \times 10^5$~K \cite{Rangelov:2016syg}, with the radius fixed to be 12~km. This is also well above the cooling theory prediction.

We also have examples of old warm ordinary pulsars: PSR J0108-1431 and B0950+08. PSR J0108-1431 is an old NS with the spin-down age of $2.0 \times 10^8$~years 
\cite{2009ApJ...701.1243D}, where the Shklovskii correction 
is taken into account. The analysis in Ref.~\cite{Mignani:2008jr} with a  Rayleigh-Jeans spectrum fit shows that its surface
temperature is $T_s^\infty = (7-10) \times 10^4 \,
(d_{130}/R_{13})^2$~K, where $d_{130}$ is the distance 
in units of 130~pc and $R_{13}$ is the apparent radius 
in units of 13~km. Within the error of the distance, 
$d = 210^{+ 90}_{-50}$~pc \cite{2012ApJ...755...39V}, the 
maximum (minimum) temperature is estimated as $T_s^\infty 
= 5.3 \times 10^5$~K ($1.1 \times 10^5$~K) for a radius 
of 13~km. PSR B0950+08 has the spin-down age of $1.75 \times 10^{7}$~years. Its surface temperature is obtained with a power-low plus blackbody spectrum fit
in Ref.~\cite{Pavlov:2017eeu} as $(1-3) \times 10^5$~K, with other parameters such as the pulsar
radius varied in a plausible range.

Seven middle aged pulsars, RX J2143.0+0654, J0806.4-4123, J0420.0-5022, J1308.6+2127, J0720.4-3125, J1856.5-3754, and J1605.3+3249 are classified into the X-ray Dim Isolated Neutron Stars (XDINSs), which are also dubbed as the Magnificent Seven. 
They exhibit thermal X-ray emission without any signature of magnetospheric activity, and have a rather long spin period.
See 
Refs.~\cite{2007Ap&SS.308..181H, vanKerkwijk:2006nr, Kaplan:2008qn} for reviews of XDINSs. 
Their surface temperatures are found to be $\sim 10^6\unit{K}$, which are again higher than the prediction of the cooling theory.
The inferred dipole magnetic fields of these NSs are relatively large: $B\sim 10^{13} - 10^{14}\unit{G}$.
As we will discuss in the following sections, in these NSs, a different type of heating mechanism due to the magnetic field decay \cite{Pons:2008fd, Vigano:2013lea} may operate.

Contrary to the above examples, PSR J2144-3933 is an old ``cool'' NS. This is one of the slowest pulsars, having $P=8.51$~s, and its spin-down age is $t_{\mathrm{sd}} = 333\unit{Myr}$ with the Shklovskii correction taken into account. Assuming the blackbody spectrum, the authors in Ref.~\cite{Guillot:2019ugf} obtained an upper limit on the surface temperature of J2144-3933: $T_s^\infty < 4.2\times 10^4\unit{K}$. This is the lowest limit on the surface temperature of NSs for the moment. 

In the next section, we discuss if the minimal cooling setup with the non-equilibrium beta processes is compatible with the observed surface temperatures in Tab.~\ref{tab:psr-temp}.

\section{Results}
\label{sec:results}

Now we show the results of our numerical analysis, where we follow the thermal evolution of NSs with the effect of the non-equilibrium beta reactions included. We then compare the results with the observed surface temperatures given in Tab.~\ref{tab:psr-temp}.

\subsection{Physical input}
\label{sec:phys-input}

We perform the numerical study in the framework of the minimal cooling with the non-equilibrium beta process discussed in Sec.~\ref{sec:therm-evol-neutr}. The following inputs are common to all of the analyses: 
\begin{itemize}
\item
  APR EOS~\cite{Akmal:1998cf}. 
\item
 Initial condition: $T^\infty = 10^{10}\unit{K}$ and $\eta_e^\infty =\eta_\mu^\infty =0$. 
\item
  Protons and neutrons form singlet ($^1S_0$) and triplet ($^3P_2\,(m_J=0)$) pairings in the core, respectively. 
\item
  The pulsar braking index $n=3$, i.e., $\Omega(t)$ obeys Eq.~\eqref{eq:P-Pdot-dipole}.
\item
  We use the formula given in Ref.~\cite{Potekhin:1997mn} for the relation between $T^\infty$ and $T_s$.
\end{itemize}
For the superfluid gap models, we use the CCDK and AO models for proton and the ``a'', ``b'' and ``c'' models for neutron, which are shown in Fig.~\ref{fig:gap}. The values of $Z_{np}$, $Z_{np\ell}$, and $W_{np\ell}$ in Eqs.~\eqref{eq:roto-diff-eta-e} and \eqref{eq:roto-diff-eta-mu} are read from Fig.~3 in Ref.~\cite{Fernandez:2005cg}, which are summarized in Tab.~\ref{tab:const}.
The numerical values of EOS and the solution of Tolman- Oppenheimer-Volkoff (TOV) equation are taken from \texttt{NSCool}~\cite{NSCool}.

\begin{table}
  \centering
  \begin{tabular}{lccccc}\toprule
    $M$ &$Z_{npe}$&$Z_{np\mu}$&$Z_{np}$&$W_{npe}$&$W_{np\mu}$ \\\relax
    [$M_\odot$] &[$10^{-61}\unit{erg}$]&[$10^{-61}\unit{erg}$]&[$10^{-61}\unit{erg}$]&[$10^{-13}\unit{erg}\unit{s^2}$]&[$10^{-13}\unit{erg}\unit{s^2}$]\\ \midrule
    $1.4$ & $10$ & $12$ & $4$ & $-1.5$ & $-2$ \\
    $1.8$ & $6$ & $7$ & $2$ & $-1.4$ & $-1.8$ \\\bottomrule
  \end{tabular}
  \caption{The values of $Z_{np}$, $Z_{np\ell}$, and $W_{np\ell}$ in Eqs.~\eqref{eq:roto-diff-eta-e} and \eqref{eq:roto-diff-eta-mu}, which are taken from Ref.~\cite{Fernandez:2005cg}.}
  \label{tab:const}
\end{table}

We divide the NSs listed in Tab.~\ref{tab:psr-temp} into two categories: MSPs and the others.
The latter contains ordinary pulsars and XDINSs. We exploit a representative parameter set for each category as follows:

\paragraph{Millisecond pulsars}

MSPs have much smaller $P$ and $\dot P$ than ordinary pulsars.
With MSP J0437-4715 in mind, we use the following parameters for this category: 
\begin{itemize}
\item
  $M = 1.4\,M_\odot$.
\item
  $P=5.8\unit{ms}$.
\item
  $\dot P = 5.7\times10^{-20}$.
\item
  $\Delta M/M = 10^{-7}$.
\end{itemize}
We also note that the values of $P$ and $\dot{P}$ of J2124-3358, $P = 4.9\unit{ms}$ and $\dot{P} = 2.1 \times 10^{-20}$, are fairly close to those of J0437-4715, while its mass is unknown. We have fixed the amount of the light elements in the envelope, $\Delta M/M = 10^{-7}$, as it turns out that the result is almost independent of this choice for old NSs such as J0437-4715 and J2124-3358.

\paragraph{Ordinary pulsars and XDINSs}

For ordinary pulsars and XDINSs, we use
\begin{itemize}
  \item
$M = 1.4\,M_\odot$ or $1.8\,M_\odot$.
\item
  $P=1\unit{s}$.
\item
  $\dot P = 1\times10^{-15}$.
\item
  $\Delta M/M = 10^{-7}$ or $10^{-15}$.
\end{itemize}
Note that $P$ and $\dot{P}$ affect the rotochemical heating only through Eq.~\eqref{eq:P-Pdot-dipole}, and thus the result depends only on the combination $P\dot P$.
Ordinary pulsars have $P\dot P \sim 10^{-17} - 10^{-13}$, corresponding to $B\sim 10^{11} - 10^{13}\unit{G}$. The dependence of the thermal evolution on $P\dot P$ is weaker than that on gap models and $P_0$, and thus we fix it to be $P\dot P = 1\times10^{-15} \unit{s}$ in the following analysis.

\vspace{5mm}
Once we fix the NS parameters as above, the time evolution of the NS surface temperature depends only on the nucleon gap models and the initial period $P_0$. As we see in Sec.~\ref{sec:phase-space-factors}, the heating rate depends on the nucleon pairing gaps via the phase space factors. On the other hand, the initial period $P_0$ affects the time evolution of the NS angular velocity $\Omega(t)$ in Eq.~\eqref{eq:P-Pdot-dipole}, through which the accumulation rate of the chemical imbalance is modified. We will study these effect in the following subsections.

\subsection{Millisecond pulsars}
\label{sec:millisecond-pulsars}

\begin{figure}
  \centering
  \begin{minipage}{0.5\linewidth}
    \includegraphics[width=1.0\linewidth]{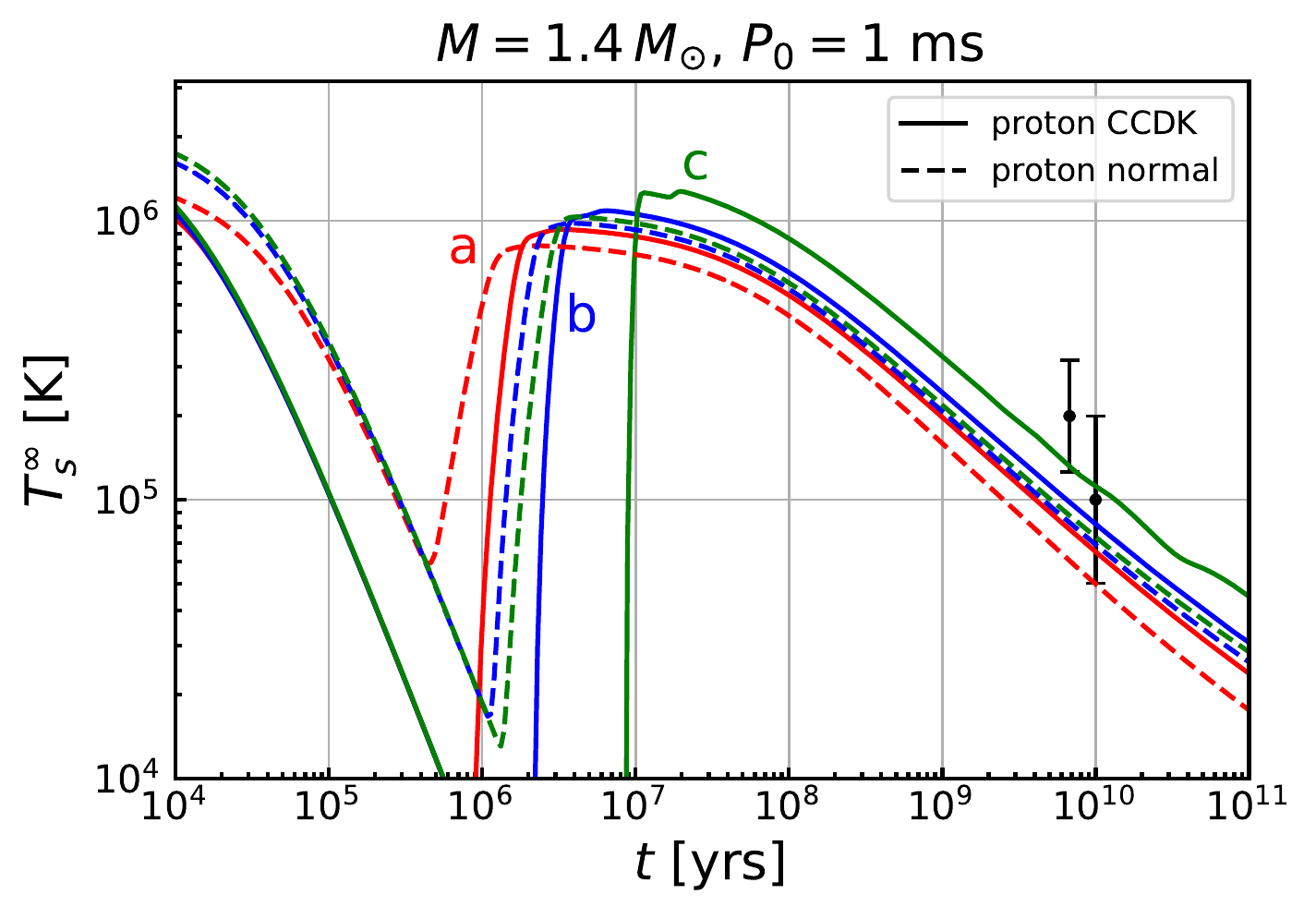}
  \end{minipage}%
  \begin{minipage}{0.5\linewidth}
    \includegraphics[width=1.0\linewidth]{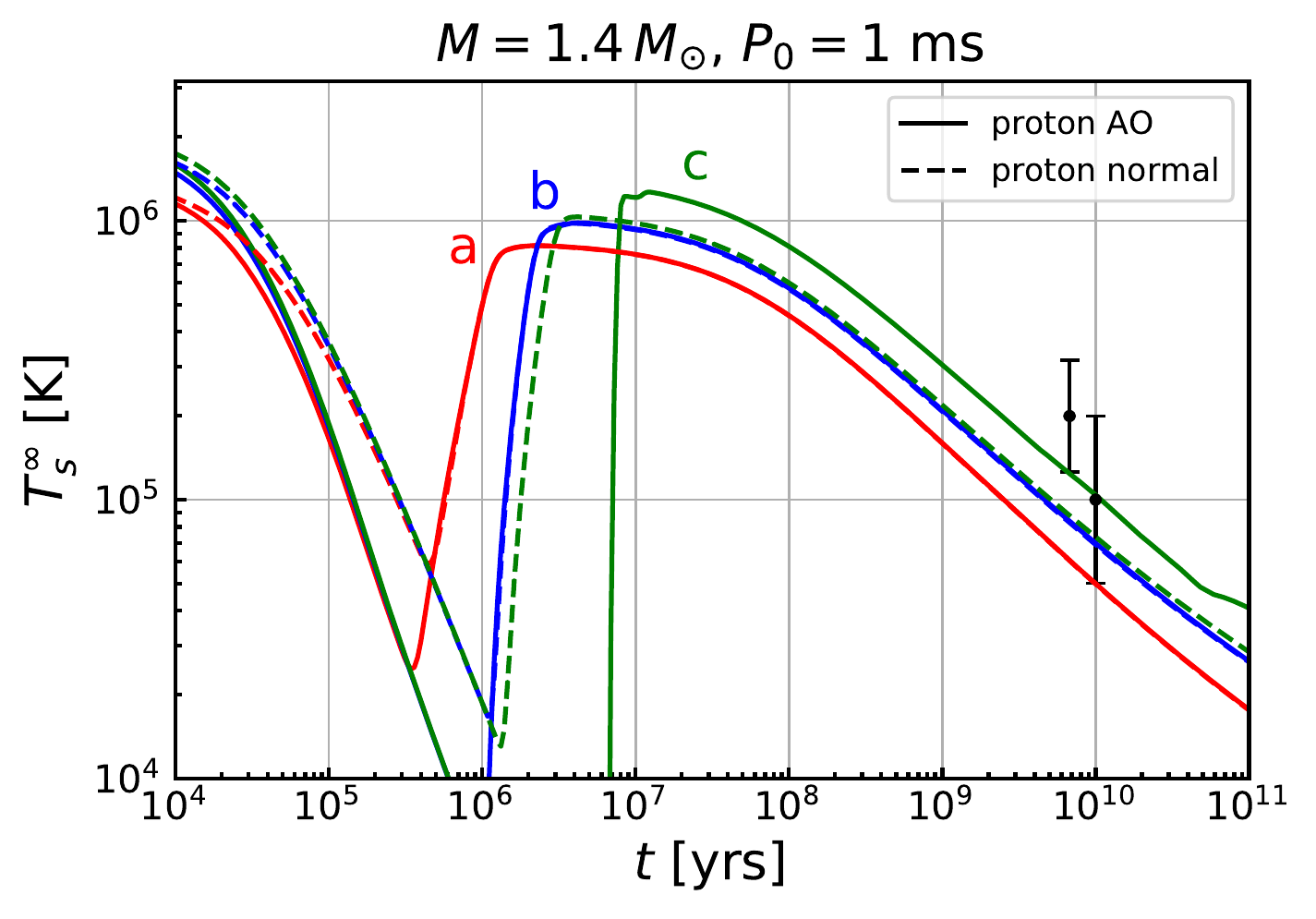}
  \end{minipage}
  \caption{Time evolution of the redshifted surface temperature $T_s^\infty$ for the MSP category with $P_0 = 1\unit{ms}$. We use the CCDK (AO) model for proton pairing in the left (right) panel. The red, blue, and green lines correspond to the ``a'', ``b'', and ``c'' models for neutron pairing, respectively. The solid (dashed) lines are for the case with (without) proton superfluidity. The observed surface temperatures of J0437-4715 and J2124-3358 are also shown by the black points with black solid lines indicating the uncertainty.}
  \label{fig:msp}
\end{figure}

We first compute the evolution of the redshifted surface temperature $T_s^\infty$ for the MSP category. The resultant temperature evolution is shown in Fig.~\ref{fig:msp}, where the initial period is taken to be $P_0=1\unit{ms}$. We use the CCDK (AO) model for proton pairing in the left (right) panel. The red, blue, and green lines correspond to the ``a'', ``b'', and ``c'' models for neutron pairing, respectively. The solid (dashed) lines are for the case with (without) proton superfluidity. The observed surface temperatures of J0437-4715 and J2124-3358 are also shown by the black points with black solid lines indicating the uncertainty.%

In all the cases given in Fig.~\ref{fig:msp}, the chemical imbalance $\eta_\ell^\infty$ overcomes the threshold of rotochemical heating $\Delta_{\mathrm{th}}$ at $t=10^{6-7}\unit{yr}$, at which the surface temperatures quickly rise to $T_s^\infty \sim 10^6\unit{K}$. After that the system reaches a quasi-steady state, where the increase in $\eta_\ell^\infty$ due to spin-down is compensated by the consumption via the Urca processes and the heating rate balances with the photon cooling rate. During this stage, $T_s^\infty$ gradually decreases due to the decline of the term $|\Omega\dot\Omega|$.

As we discussed in Sec.~\ref{sec:therm-evol-neutr}, in this work we include the effect of both proton and neutron superfluidity simultaneously. We can see the consequence of this simultaneous inclusion by comparing the solid and dashed lines for each case. Let us first study the case with the proton CCDK for $P_0 = 1\unit{ms}$, i.e., the left panel in Fig.~\ref{fig:msp}. In this case, for all neutron gap models, the heating effect starts to be visible at a later time in the presence of proton superconductivity. This is because the additional contribution from the proton pairing gap increases the threshold of rotochemical heating $\Delta_{\mathrm{th}}$, and thus an extra amount of $\eta_\ell^\infty$ needs to be accumulated. This results in a delay in the onset of rotochemical heating. Moreover, a larger value of $\Delta_{\mathrm{th}}$ leads to a larger value of $\eta_\ell^\infty$ eventually. This then results in a higher $T_s^\infty$ at late times, since the heating power is proportional to $\eta_\ell$ as in Eq.~\eqref{eq:hating-rate}. This feature can also be seen for every neutron gap model in the left panel in Fig.~\ref{fig:msp}.

Next, we examine the cases with the AO proton gap model shown in the right panel in Fig.~\ref{fig:msp}. For this proton paring gap, we do not see enhancement in $T_s^\infty$ due to the proton superfluidity for the neutron gaps ``a'' and ``b''. As we see in the left panel in Fig.~\ref{fig:gap}, the size of the AO proton gap is smaller than the CCDK gap, and even vanishes deep inside the NS core. For this reason, the rotochemical threshold $\Delta_{\mathrm{th}}$ is determined almost solely by the neutron gap, which makes the effect of proton superconductivity invisible. 
For the neutron ``c'' gap, on the other hand, the gap amplitude is very large near the NS center, and hence heating is ineffective there. Instead, the rotochemical heating mainly occurs in the intermediate region where the AO proton gap is sizable, which makes the effect of proton gap manifest. This observation indicates that it is crucial to take account of the density dependence of the nucleon pairing gaps for the evaluation of the rotochemical heating effect.

The results in Fig.~\ref{fig:msp} show that the observed surface temperatures of J0437-4715 and J2124-3358 can be explained by the heating effect of non-equilibrium beta reactions, especially for  moderate/large nucleon gaps. In particular, for the neutron ``c'' gap model, the simultaneous inclusion of proton superconductivity improves the fit considerably such that the predicted thermal evolution is totally consistent with the observed temperatures.

\subsection{Ordinary pulsars and XDINSs}
\label{sec:ordinary-pulsars}

\begin{figure}
  \centering
  \begin{minipage}{0.5\linewidth}
    \includegraphics[width=1.0\linewidth]{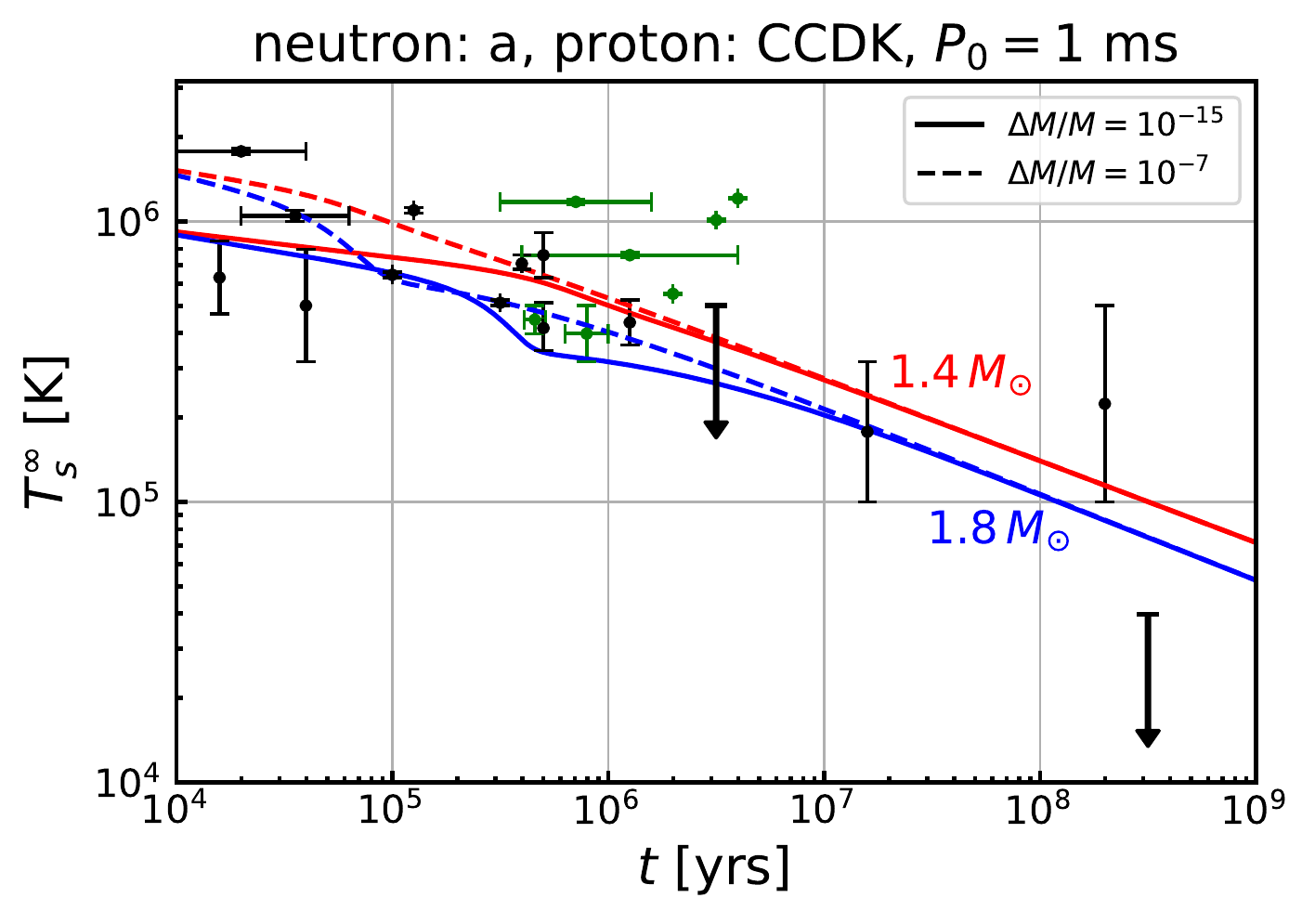}
  \end{minipage}%
  \begin{minipage}{0.5\linewidth}
    \includegraphics[width=1.0\linewidth]{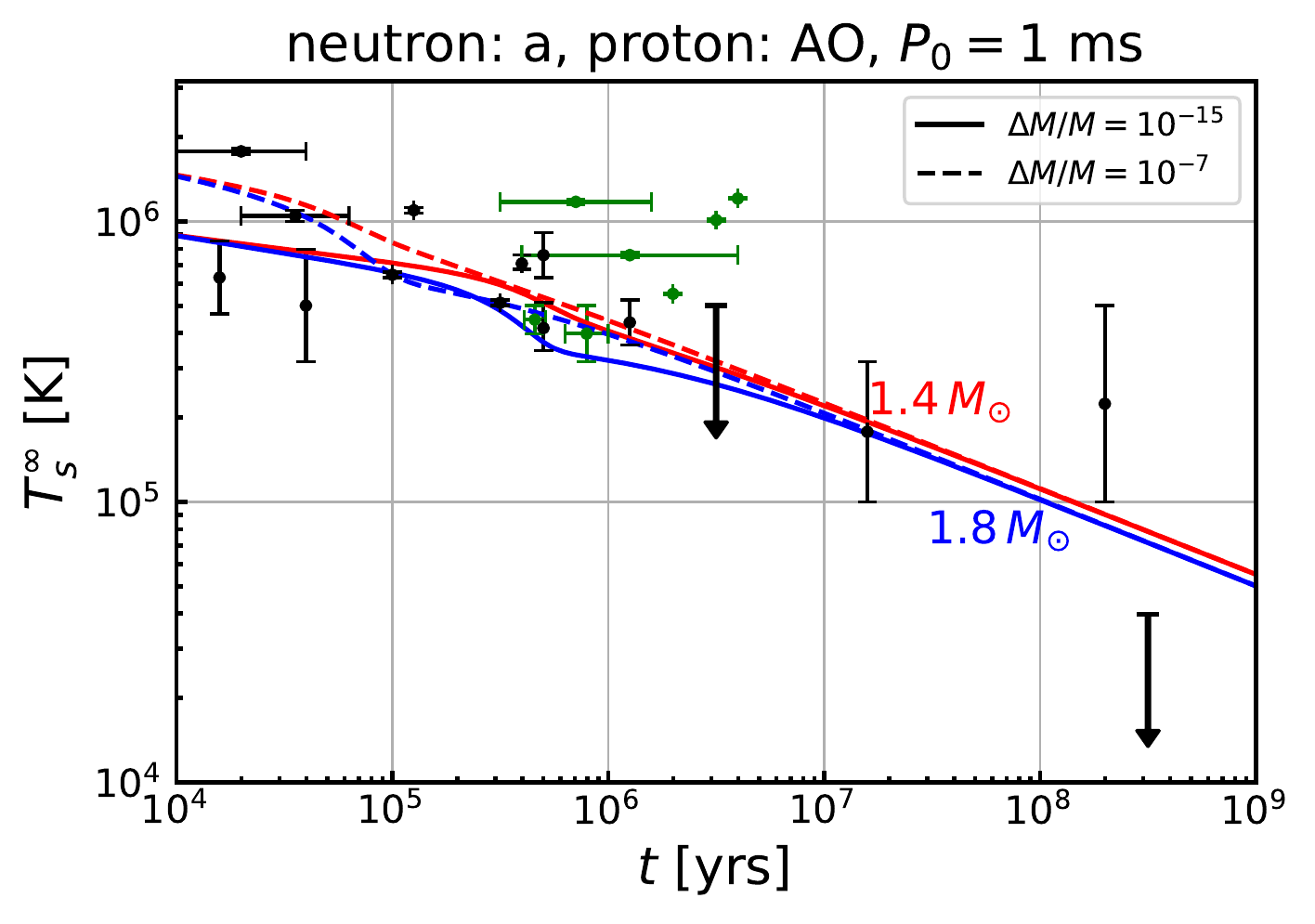}
  \end{minipage}
  \begin{minipage}{0.5\linewidth}
    \includegraphics[width=1.0\linewidth]{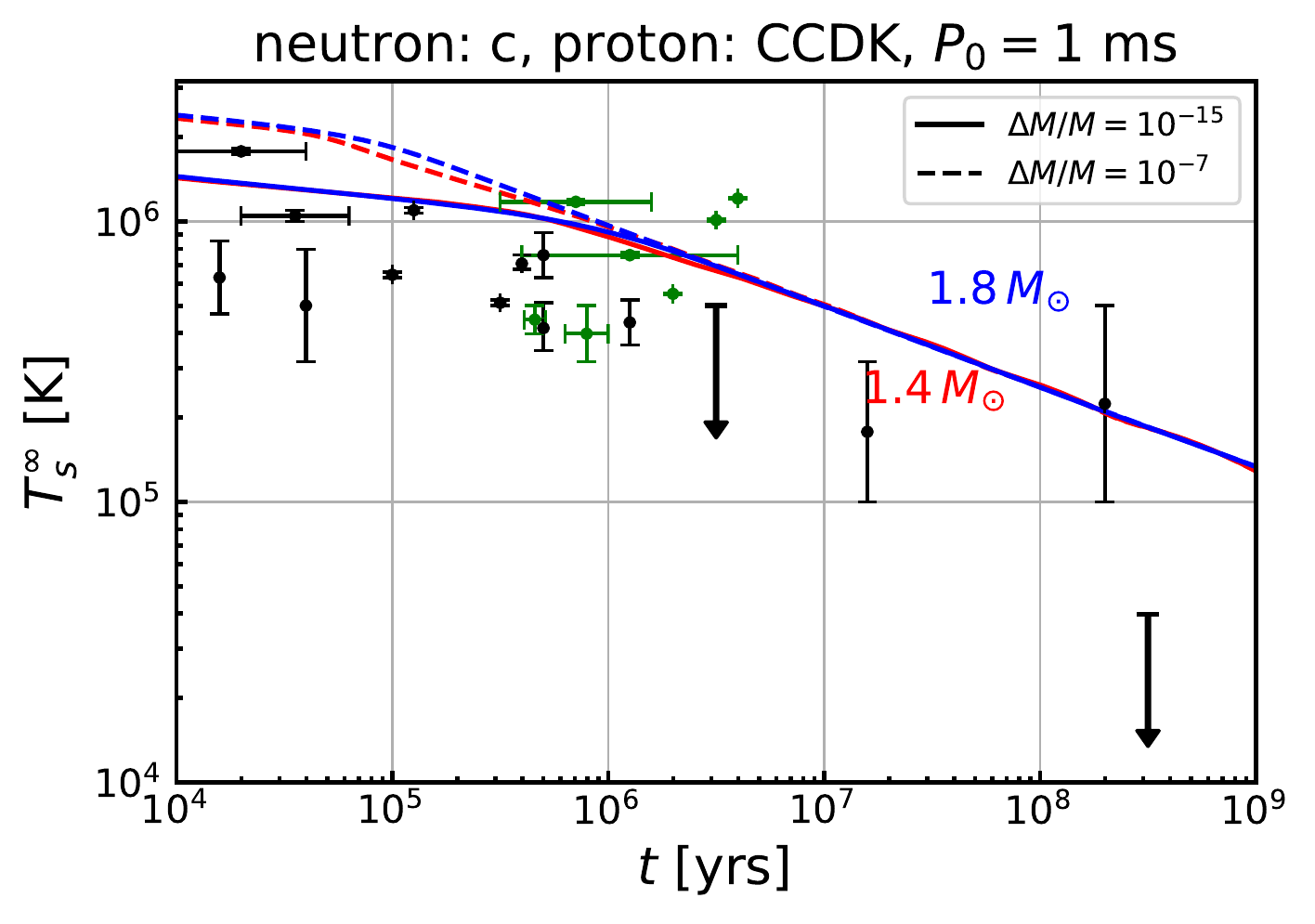}
  \end{minipage}%
  \begin{minipage}{0.5\linewidth}
    \includegraphics[width=1.0\linewidth]{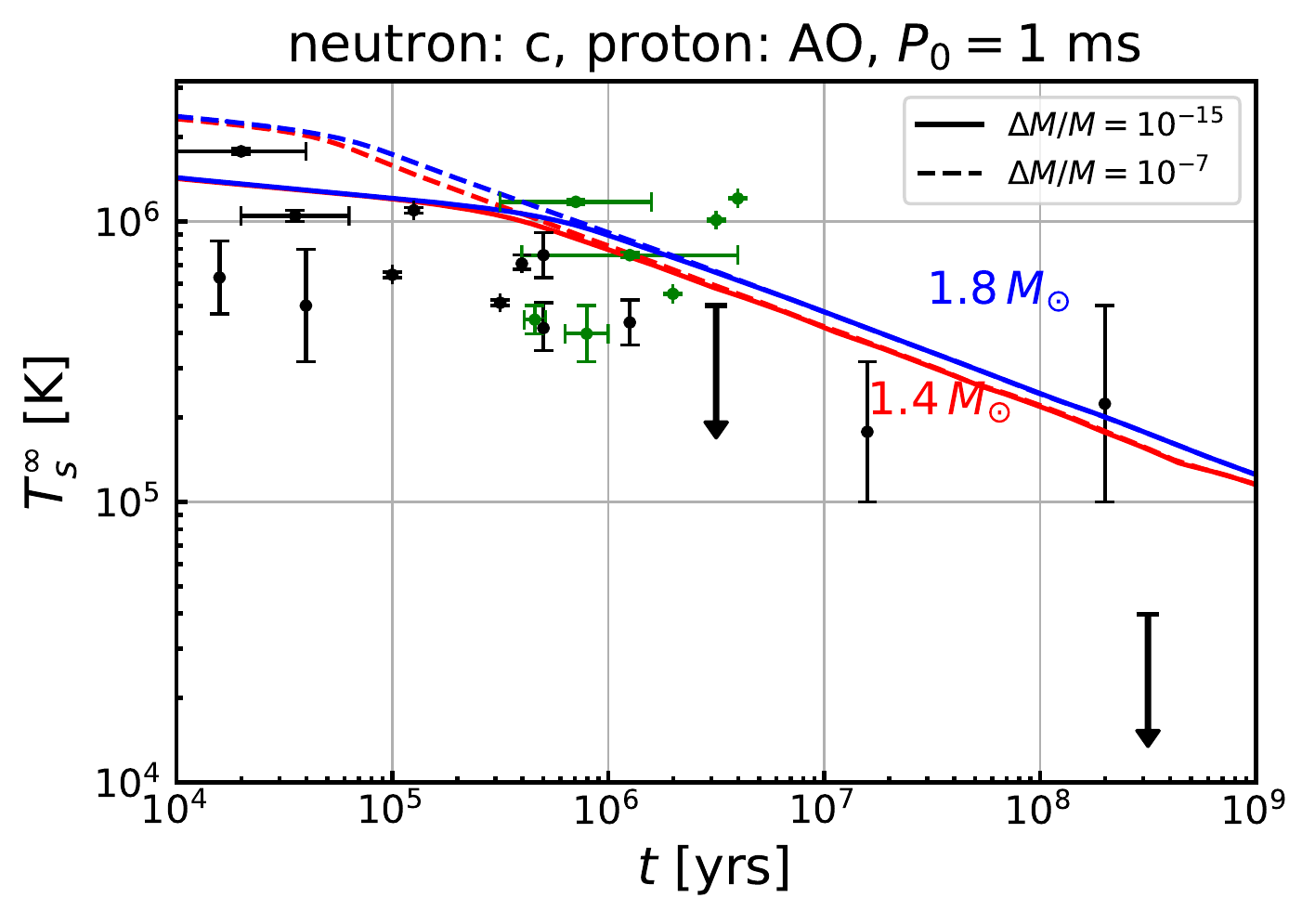}
  \end{minipage}
  \caption{Time evolution of the surface temperature $T_s^\infty$ for ordinary pulsars and XDINSs with $P_0=1\unit{ms}$. We use the neutron gap ``a'' (``c'') in the upper (lower) panels and the CCDK (AO) proton gap in the left (right) panels. The red and blue lines correspond to $M=1.4\,M_\odot$ and $1.8\,M_\odot$, while the solid and dashed lines represent the cases for the heavy ($\Delta M/M = 10^{-15}$) and light ($\Delta M/M = 10^{-7}$) element envelope models, respectively. We also plot the observed surface temperatures of ordinary pulsars and XDINSs with black and green points, respectively, with the horizontal (vertical) lines indicating the uncertainty in the kinematic age ($T_s^\infty$).}
  \label{fig:cp-1}
\end{figure}

Next we discuss the second category, which is comprised of ordinary pulsars and XDINSs. In Fig.~\ref{fig:cp-1}, we show the time evolution of the surface temperature $T_s^\infty$ for this category with $P_0=1\unit{ms}$. We use the neutron gap ``a'' (``c'') in the upper (lower) panels and the CCDK (AO) proton gap in the left (right) panels. The red and blue lines correspond to $M=1.4\,M_\odot$ and $1.8\,M_\odot$, while the solid and dashed lines represent the cases for the heavy ($\Delta M/M = 10^{-15}$) and light ($\Delta M/M = 10^{-7}$) element envelope models, respectively. It is found that another choice of the envelope parameter $\Delta M/M$ just falls in between these two cases.
We also plot the observed surface temperatures of ordinary pulsars and XDINSs with black and green points, respectively, with the horizontal (vertical) lines indicating the uncertainty in the kinematic age ($T_s^\infty$). The bars with down arrows correspond to the upper limits on $T_s^\infty$ of J2144-3933 and B1929+10.

For the cases shown in Fig.~\ref{fig:cp-1}, the rotochemical heating begins earlier than $10^4\unit{yr}$.
The difference from the MSP category is due to the larger $P\dot P$ of the ordinary pulsars and XDINSs. 
The predicted temperatures tend to be higher for the light element envelope than the heavy element one at earlier times, but this difference disappears at later times. In addition, the evolution curves depend on the NS mass. This dependence is, however, rather non-trivial compared with that observed in Ref.~\cite{Gonzalez-Jimenez:2014iia}, where only the neutron triplet paring is taken into account and the temperature always gets higher for a lighter NS mass. This complexity is caused by the non-trivial dependence of the rotochemical threshold on the matter density, as both proton and neutron pairing gaps contribute to the threshold. 

It is found from the top two panels that the small neutron gap ``a'' offers the predictions consistent with most of the surface temperatures of the ordinary pulsars. 
We also find that three hot XDINSs, having relatively high temperatures $T_s^\infty \sim 10^6\unit{K}$ at $t\sim 10^6\unit{yr}$, are located above the predictions.
The bottom two panels, on the other hand, show that the large neutron gap ``c'' gives higher temperatures than the ``a'' model, and can be consistent only with a part of the NS temperatures shown in the figure. The hot XDINSs are still above the prediction, while several young pulsars ($t\sim 10^4\unit{yr}$ and $T_s^\infty < 10^6\unit{K}$) are below the thermal evolution curves.

\begin{figure}
  \centering
  \begin{minipage}{0.5\linewidth}
    \includegraphics[width=1.0\linewidth]{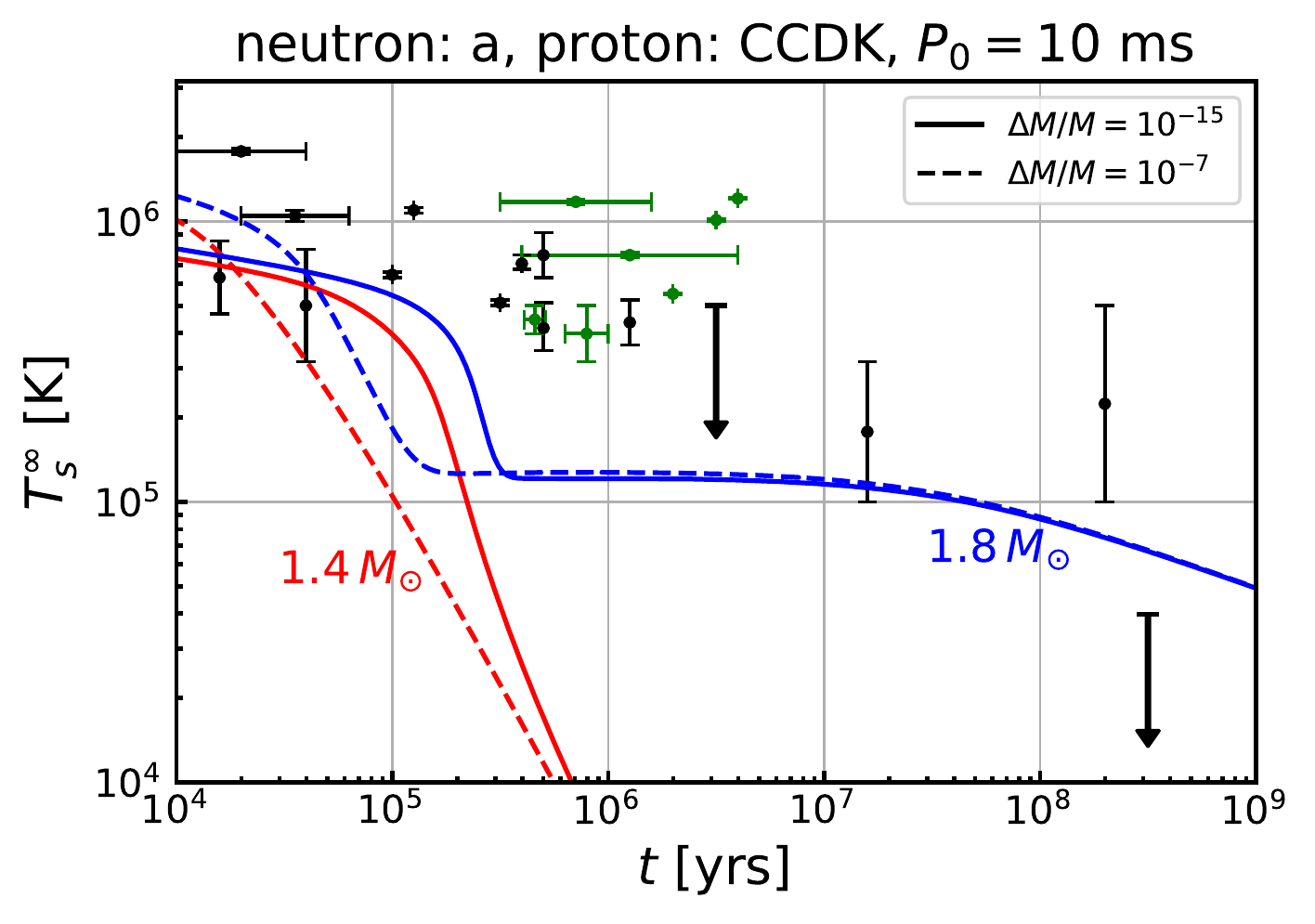}
  \end{minipage}%
  \begin{minipage}{0.5\linewidth}
    \includegraphics[width=1.0\linewidth]{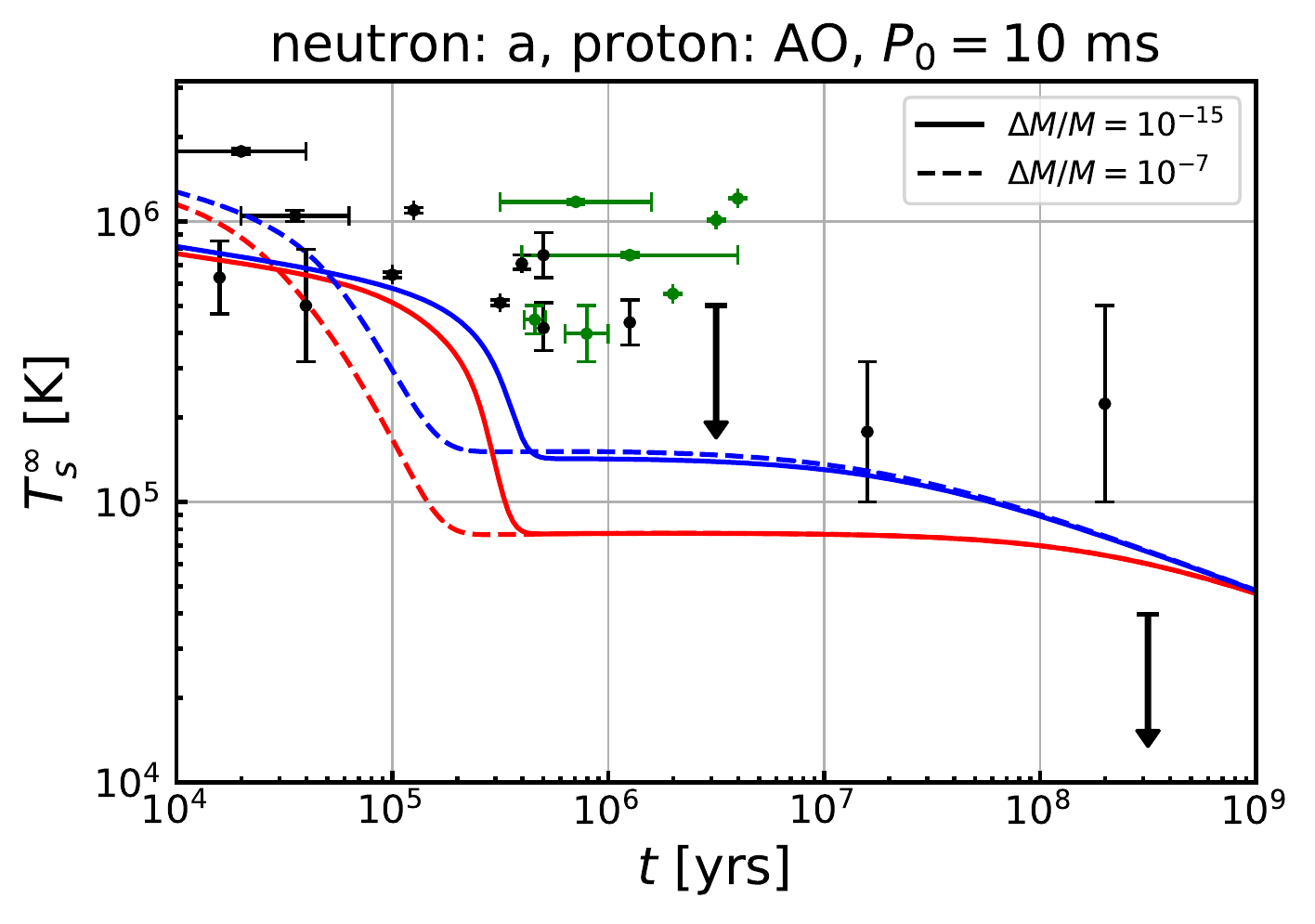}
  \end{minipage}
  \begin{minipage}{0.5\linewidth}
    \includegraphics[width=1.0\linewidth]{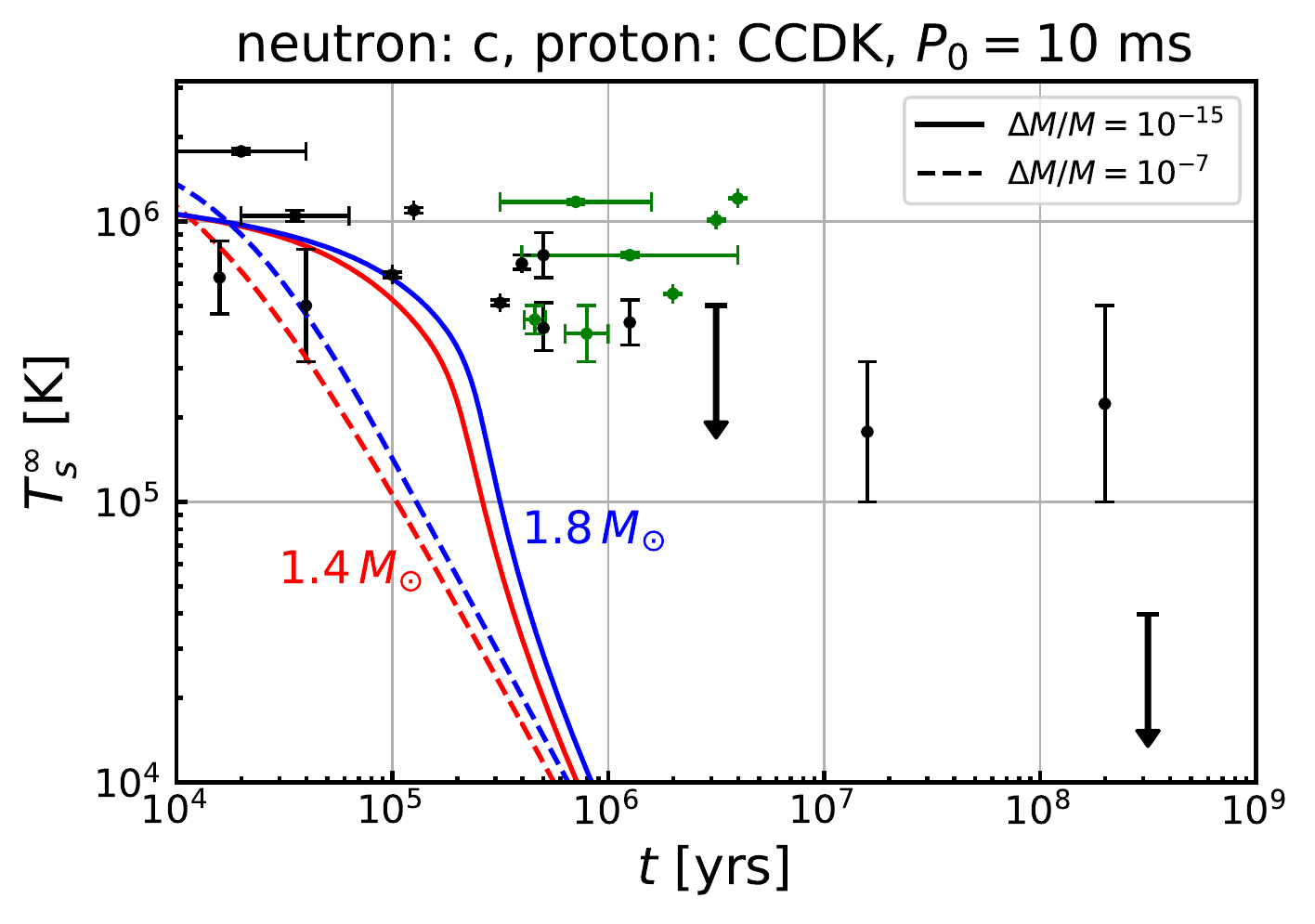}
  \end{minipage}%
  \begin{minipage}{0.5\linewidth}
    \includegraphics[width=1.0\linewidth]{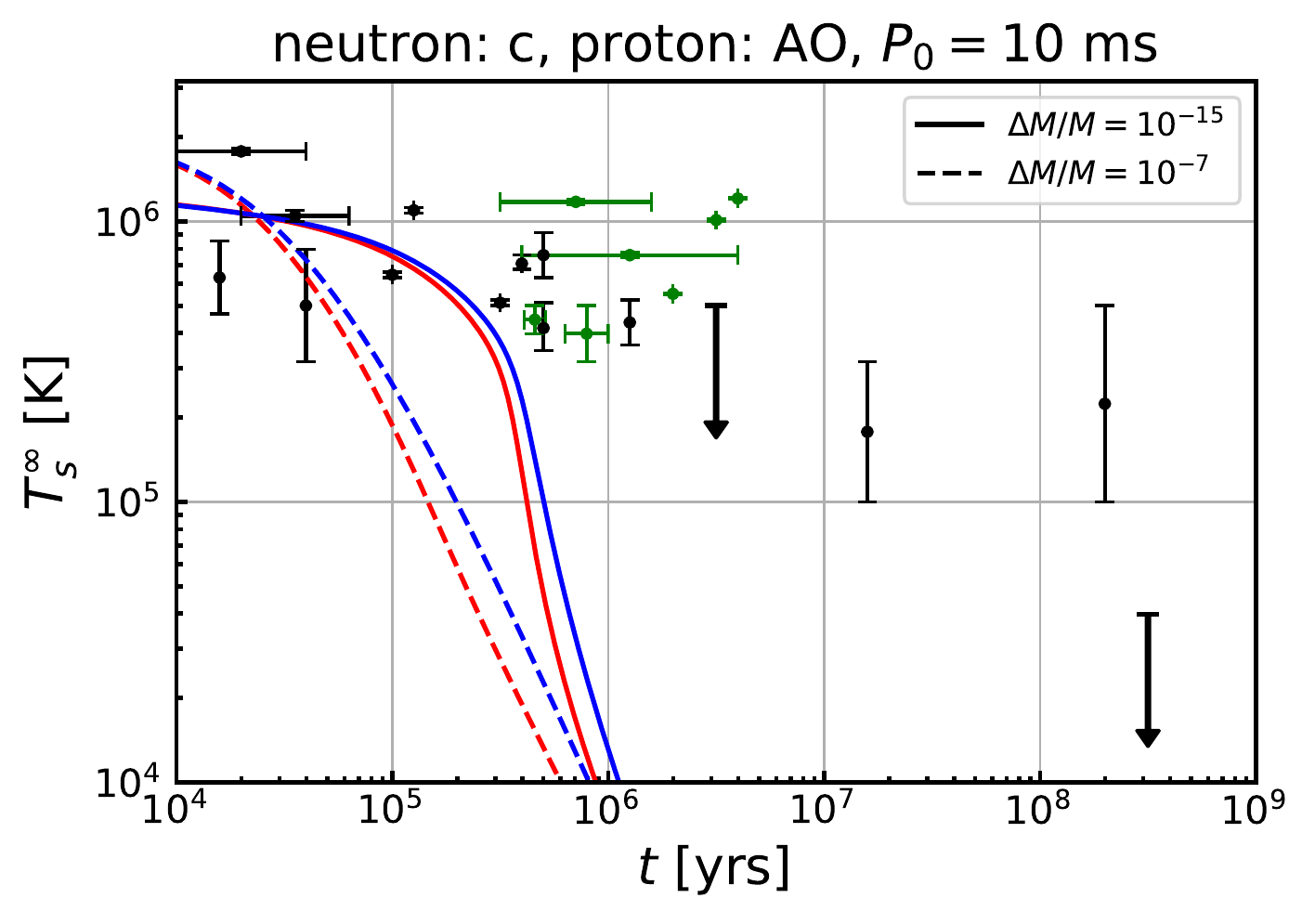}
  \end{minipage}
  \caption{The same as the left panels in Fig.~\ref{fig:cp-1} for $P_0=10\unit{ms}$, where the meaning of the lines is the same as in the figure. }
  \label{fig:cp-2}
\end{figure}

A different choice of $P_0$ may change the evolution of $T_s^\infty$ via the modification of the increase rate in $\eta_\ell^\infty$. In general, a larger initial period makes the departure from the beta equilibrium smaller and hence the heating effect milder. To see this, in Fig.~\ref{fig:cp-2}, we show the time evolution of the surface temperature for $P_0=10\unit{ms}$ for the CCDK (left) and AO (right) proton gap models and the ``a'' (top) or ``c'' (bottom) neutron gap models. It is found that this slightly larger initial period strongly suppresses the heating effect. 
This can be seen in the bottom panels, where the heating does not occur for both of the NS masses due to the large neutron gap of ``c''. For the neutron gap ``a'', on the other hand, the internal heating can occur when the nucleon pairing gaps are sufficiently suppressed in the NS center so that even a small amount of the chemical imbalance can trigger the rotochemical heating. The predictions in all of these cases are consistent with the surface temperatures of relatively young pulsars ($t\sim 10^4 - 10^6\unit{yr}$) of $10^5\unit{K}<T_s^\infty < 10^6\unit{K}$. The surface temperature of B0950+08 may be explained for some cases. 
The upper limit on the surface temperature of J2144-3933 can also be satisfied.

We note that, for most of the NS parameter choices, at later times, the intensity of the rotochemical heating has little dependence on $P\dot{P}$ and thus on its dipole magnetic field, as discussed in detail in App.~\ref{sec:magnetic-field}. This is because for a sufficiently old pulsar, the spin-down effect is very weak and the evolution of $\eta_\ell^\infty$ is solely determined by the equilibration caused by the modified Urca process. 
This observation, therefore, indicates that we cannot explain the low surface temperature of J2144-3933 by changing the magnetic field while keeping $P_0$ small. 

In summary, the observed surface temperatures of ordinary pulsars are compatible with the rotochemical heating for both small and large neutron gaps. For a small neutron gap, the predicted curves of the time evolution of $T_s^\infty$ tend to be in good agreement with most of the observations for $P_0 = 1\unit{ms}$. In the case of a large neutron gap, on the other hand, the middle-aged and old pulsars can still be explained with $P_0 = 1\unit{ms}$, while it is required to assume a larger initial period for young pulsars; for an initial period of $1\unit{ms} < P_0 < 10\unit{ms}$, the heating can still occur but predicted temperatures tend to be lower than those for $P_0 = 1\unit{ms}$, with which we can explain all of the data below the curves shown in Fig.~\ref{fig:cp-1}. In particular, the J2144-3933 limit can be satisfied for any nucleon gap models if we take a sufficiently large initial period. It is intriguing to note that the J2144-3933 has one of the longest periods observed so far, as mentioned in Sec.~\ref{sec:observations-old-hot}. This observation may suggest that the initial period of this NS is also relatively long so that the rotochemical heating has never operated and the NS has been cooled down below the limit $T_s^\infty < 4.2\times 10^4\unit{K}$.

On the other hand, the observed temperatures of some of the XDINSs may be higher than the prediction of the rotochemical heating. This discrepancy could be merely due to the errors of their age and temperature; in particular, the temperatures shown in the Tab.~\ref{tab:psr-temp} may suffer from large systematic uncertainty because of our ignorance of their masses, radii, and distances. We also note that the temperature data of the three especially hot XDINS---J2143+0654, RX J0806.4-4123, and RX J1308.6+2127---used in this analysis are obtained using a spectrum fit with one blackbody component plus Gaussian absorption lines. If there is a hot spot in these NSs, however, such a fit tends to give a higher temperature and a smaller NS radius than the actual ones. In fact, the inferred radii of J2143+0654 and RX J0806.4-4123, $R = 3.10(4)\unit{km}$ for $d = 500\unit{pc}$ \cite{Kaplan:2009au} and $R=2.39(15)\unit{km}$ for $d = 250\unit{pc}$ \cite{Kaplan:2009ce}, respectively, are considerably smaller than the typical size of the NS radius $\sim 10-15\unit{km}$. In addition, these NSs exhibit X-ray pulsations \cite{Pires:2014qza}, which indicate that the temperature distribution on their surface is inhomogeneous. It is indeed found that a spectrum analysis based on a two-temperature blackbody model gives a larger radius and a lower surface temperature for these XDINSs \cite{Yoneyama:2018dnh}, with which the discrepancy between the prediction and observation is significantly reduced. 
Another potential explanation for the discrepancy is that the XDINSs have undergone the magnetic field decay, which makes the spin-down age differ from the actual age. 
Moreover, the magnetic field decay itself can be another source of heating~\cite{2012MNRAS.422.2878D, Vigano:2013lea}. If this is the case, the inclusion of only the rotochemical heating effect may be insufficient to explain the temperatures of the XDINSs.

\section{Conclusions}
\label{sec:conclusions}

We have studied the non-equilibrium beta process in the minimal cooling scenario, which gives rise to the late time heating in NSs. Extending the previous works, we have included the singlet proton and triplet neutron pairing gaps simultaneously in the calculation of the rate and emissivity of the process, with their density dependence taken into account. We then compare the time evolution of the NS surface temperature predicted in this framework with the latest observations of the NS temperatures, especially with the recent data of the old warm NSs. It is found that the simultaneous inclusion of both proton and neutron gaps is advantageous for the explanation of the old warm NSs, since it increases the threshold of rotochemical heating and thus enhances the heating effect. We find that the observed surface temperatures of warm MSPs, J2124-3358 and J0437-4715, are explained for various choices of nucleon gap models. The same setup can also explain the temperatures of ordinary pulsars by choosing the initial rotational period of each NS accordingly. In particular, with $P_0 = 10\unit{ms}$ or larger, the upper limit on the surface temperature of J2144-3933 can be satisfied.

To explain the observation, we require $P_0\lesssim 10\unit{ms}$ for old warm pulsars and $P_0 \gtrsim 10\unit{ms}$ for old cold one. This assumption is reasonable for hot MSPs since their current periods are also as small as $\Order(1)\unit{ms}$. On the other hand, for ordinary pulsars, several recent studies suggest that they are born with $P_0 = \Order(10-100)\unit{ms}$~\cite{2012Ap&SS.341..457P, 2013MNRAS.430.2281N, Igoshev:2013rqf, FaucherGiguere:2005ny, 2010MNRAS.401.2675P, Gullon:2014dva, Gullon:2015zca, Muller:2018utr}, 
which is apparently in tension with the requirement of $P_0\lesssim 10\unit{ms}$. 
Nevertheless, the initial period of a NS highly depends on the detail of the supernova process where the NS was created. Given that a fully satisfactory simulation for supernova explosion process has not yet been available, we regard the issue of the NS initial period as an open question, expecting future simulations and astrophysical observations to answer this problem.

As we have seen in Sec.~\ref{sec:ordinary-pulsars}, the surface temperatures of the three XDINSs J2143+0654, RX J0806.4-4123, and RX J1308.6+2127 are higher than the prediction of the rotochemical heating. This discrepancy may be due to large magnetic fields of these XDINSs. In fact,
XDINSs commonly have absorption feature in the X-ray spectrum~\cite{Borghese:2015iqa, Borghese:2017vxn},
which is interpreted as the proton cyclotron resonance or atomic transition.
Both require the magnetic field larger than about $10^{13}\unit{G}$, and such a strong magnetic field can affect the thermal evolution of XDINSs.
If this is the case, we need a more involved analysis for these XDINSs including the magnetic field evolution.

Regarding the ambiguity in the choice of pairing gaps, there is a hint from the observation.
The recently observed rapid cooling of the NS in the supernova remnant Cassiopeia A (Cas A) may be the first observational signature of the neutron superfluidity in the NS core~\cite{Page:2010aw, 2011MNRAS.412L.108S, Ho:2014pta, Hamaguchi:2018oqw, Wijngaarden:2019tht}.
The measured decline in temperature is explained with a large proton gap such as the CCDK model and a small neutron gap with the critical temperature of $T_c \simeq 5\times 10^8\unit{K}$, which is about a factor of 2 smaller than that for the ``a'' model.
The time evolution of $T_s^\infty$ for these gaps is close to that for the proton CCDK + neutron ``a'' model, and thus is consistent with the observations of ordinary pulsars as shown in Fig.~\ref{fig:cp-1}. This setup can also explain the surface temperature of the MSP J2124-3358, while the predicted value of $T_s^\infty$ for J0437-4715 is slightly below the observed one. Although a large neutron gap such as ``c'' model is favored by the observation of these MSPs, it is not consistent with the Cas A NS observation. Further exploration of a unified explanation for the temperatures of the Cas A NS and the old warm NSs will be given on another occasion.  

Finally, we emphasize that the non-equilibrium beta process discussed in this paper is not an adhoc assumption but an inevitable consequence of rotating NSs, and therefore the heating mechanism based on this process should always be taken into account. 
In this sense, the minimal cooling plus rotochemical heating is \textit{the minimal} scenario for the NS thermal evolution. Intriguingly, we have found that this ``minimal'' setup is compatible with the observed surface temperatures of NSs for the moment, without relying on exotic physics. Further developments in the evaluation of nucleon pairing gaps, as well as additional data of NS surface temperatures, allow us to test this minimal scenario in the future.

\section*{Acknowledgements}

We thank Teruaki Enoto, Kenji Fukushima, and Hideyuki Umeda for valuable discussions and suggestions.
We also thank Mikhail E. Gusakov for useful correspondence.
This work is supported in part by the Grant-in-Aid for
Scientific Research A (No.16H02189 [KH]),
Young Scientists B (No.17K14270 [NN]), Innovative Areas
(No.26104001 [KH], No.26104009 [KH], No.18H05542 [NN]).
The work of KY was supported by JSPS KAKENHI Grant Number JP18J10202.


\section*{Appendix}
\appendix
\section{Phase space factors}
\label{sec:phase-space}

In this appendix, we give a detailed discussion on the phase space integrals in Eqs.~\eqref{eq:i-integ-emis}  and~\eqref{eq:i-integ-gamma}. While the nucleon pairing is negligible, we use the formulas in Ref.~\cite{Reisenegger:1994be}. In the numerical calculation, we neglect the pairing if $3v_n+v_p < 1$ for the neutron branch and if $v_n+3v_p < 1$ for the proton branch.
If the pairing operates but $\xi_\ell < 1$, we neglect non-equilibrium effects. We use the results in Ref.~\cite{Gusakov:2002hh} for $I^N_{M,\epsilon}$, and set $I^N_{M,\Gamma}=0$. If the pairing operates and $\xi_\ell \geq 1$, we numerically perform the integrals using the zero temperature approximation; we show the detail of this calculation in this appendix. 
We have checked that the result scarcely depends on the choice of the threshold value of pairing and $\xi_\ell$.

\subsection{Energy integral}
\label{sec:energy-integral}

We define the energy integral part of the phase space factors by\footnote{$\tilde I^N_{M,\Gamma}$ in Eq.~\eqref{eq:itildenmgamdef} has the opposite sign to $I^N_{M,\Gamma}$ in Ref.~\cite{Petrovich:2009yh}. }
\begin{align}
  \tilde I^N_{M,\epsilon}
  &= \frac{60480}{11513\pi^8}\int_0^\infty dx_\nu \int_{-\infty}^{\infty} dx_ndx_pdx_{N_1}dx_{N_2} x_\nu^3
    f(z_n)f(z_p)f(z_{N_1})f(z_{N_2})\notag\\
  &\times\left[f(x_\nu - \xi_\ell -z_n-z_p-z_{N_1}-z_{N_2})
    + f(x_\nu + \xi_\ell -z_n-z_p-z_{N_1}-z_{N_2})\right]\,,
  \\
  \tilde I^N_{M,\Gamma}
  &= \frac{60480}{11513\pi^8}\int_0^\infty dx_\nu \int_{-\infty}^{\infty}dx_ndx_pdx_{N_1}dx_{N_2} x_\nu^2
    f(z_n)f(z_p)f(z_{N_1})f(z_{N_2})\notag\\
  &\times\left[f(x_\nu - \xi_\ell -z_n-z_p-z_{N_1}-z_{N_2})
    - f(x_\nu + \xi_\ell -z_n-z_p-z_{N_1}-z_{N_2})\right]\,.
    \label{eq:itildenmgamdef}
\end{align}
Following the argument in Ref.~\cite{Petrovich:2009yh}, we use the zero temperature approximation in this calculation; namely, 
we replace the Fermi-Dirac distribution $f(x)$ by the step function $\Theta(-x)$. The phase space factors are then written in terms of the following integral expressions:
\begin{align}
  \tilde I^N_{M,\epsilon}
  &=
    \frac{60480}{11513\pi^8}
    \frac{\xi_\ell^8}{4}
    \Theta(1-r_n-r_p-r_{N_1}-r_{N_2})
    \notag\\
    &\times
    \int_{r_n}^{1-r_{N_1}-r_{N_2}-r_p}du_n
    \int_{r_{N_1}}^{1-u_n-r_{N_2}-r_p}du_{N_1}
      \int_{r_{N_2}}^{1-u_n-u_{N_1}-r_p}du_{N_2} 
      \notag\\
  &\times
  \frac{u_{n}}{\sqrt{u_{n}^2-r_{n}^2}}
  \frac{u_{N_1}}{\sqrt{u_{N_1}^2-r_{N_1}^2}}
      \frac{u_{N_2}}{\sqrt{u_{N_2}^2-r_{N_2}^2}}
    K_\epsilon(u_n+u_{N_1}+u_{N_2}, r_p) \, ,
    \label{eq:itildenmep}
\\
  \tilde I^N_{M,\Gamma}
  &=
    \frac{60480}{11513\pi^8}
    \frac{\xi_\ell^7}{3}
    \Theta(1-r_n-r_p-r_{N_1}-r_{N_2})
    \notag\\
    &\times
    \int_{r_n}^{1-r_{N_1}-r_{N_2}-r_p}du_n
    \int_{r_{N_1}}^{1-u_n-r_{N_2}-r_p}du_{N_1}
      \int_{r_{N_2}}^{1-u_n-u_{N_1}-r_p}du_{N_2}
      \notag\\
  &\times
  \frac{u_{n}}{\sqrt{u_{n}^2-r_{n}^2}}
  \frac{u_{N_1}}{\sqrt{u_{N_1}^2-r_{N_1}^2}}
      \frac{u_{N_2}}{\sqrt{u_{N_2}^2-r_{N_2}^2}}
    K_\Gamma(u_n+u_{N_1}+u_{N_2}, r_p)\,,
    \label{eq:itildenmgam}
\end{align}
where $r_N=v_N/\xi$, $u_N=|z_N|/\xi$, and 
\begin{align}
  K_\epsilon(u,r)
  &= -\frac{1}{2}r^2(1-u)\left(3r^2 +4(1-u)^2\right)\ln\left[\frac{1-u+\sqrt{(1-u)^2-r^2}}{r}\right]
    \notag\\
  &+\frac{1}{30}\left(16r^4 + 83r^2(1-u)^2+6(1-u)^4\right)\sqrt{(1-u)^2-r^2}\,,
  \label{eq:kepdef}
  \\
  K_\Gamma(u,r)
  &= -\frac{3}{8}r^2\left(r^2 +4(1-u)^2\right)\ln\left[\frac{1-u+\sqrt{(1-u)^2-r^2}}{r}\right]
    \notag\\
  &+\frac{1}{8}(1-u)\left(13r^2 + 2(1-u)^2\right)\sqrt{(1-u)^2-r^2}\,.
\end{align}
Notice that the integral region of Eqs.~\eqref{eq:itildenmep} and \eqref{eq:itildenmgam} is different from that in Eqs.~(B.12) and (B.13) in Ref.~\cite{Petrovich:2009yh}. We also find that the sign of the first term in Eq.~\eqref{eq:kepdef} is opposite to that in Eq.~(B.11) in Ref.~\cite{Petrovich:2009yh}. 

\subsection{Angular integral}
\label{sec:angular-integral}

Next, we perform the angular integration.
Due to the different angular dependence of proton and neutron pairing gaps, we need to separately treat the integrals for the proton and neutron branches.

\paragraph{Proton branch}

For the proton branch, we can carry out the angular integration of the momenta of the protons and leptons trivially since the proton singlet gap is isotropic. The integration with respect to the neutron momentum direction is then reduced to a simple average as 
\begin{align}
  I^p_{M,\epsilon} &= \int\frac{d\Omega_n}{4\pi} \tilde I^p_{M,\epsilon}\,,
  \nonumber\\
  I^p_{M,\Gamma} &= \int\frac{d\Omega_n}{4\pi} \tilde I^p_{M,\Gamma}\,,
\end{align}
where only $v_n$ is dependent on $\Omega_n$.

\paragraph{Neutron branch}

The neutron branch involves three neutrons, $n$, $n_1$ and $n_2$, so the angular integral is generically complicated. 
Here we neglect the proton and lepton momenta in the momentum conserving delta function.
Then, the three neutron momenta form an equilateral triangle, and we can put the angular integral into
\begin{align}
  I^n_{M,\epsilon}
  &=
    \int_0^1d\cos\theta_n\int_0^{2\pi}\frac{d\varphi_{n_1}}{2\pi}
    \tilde I^n_{M,\epsilon}\,,
  \nonumber\\
  I^n_{M,\Gamma}
  &=
    \int_0^1d\cos\theta_n\int_0^{2\pi}\frac{d\varphi_{n_1}}{2\pi}
    \tilde I^n_{M,\Gamma}\,,
    \label{eq:angintn}
\end{align}
with $\cos\theta_n$ the polar angle of $n$ around the quantization axis and $\varphi_{n_1}$ the azimuthal angle of $n_1$ around $\bm p_n$.
The relative angles among the momenta of $n$, $n_1$ and $n_2$ are fixed because of the momentum conservation. The polar angles of $\bm{p}_{n_1}$ and $\bm{p}_{n_2}$ with respect to the quantization axis are written as
\begin{align}
  \cos\theta_{n_1}
  &=
    -\frac{\sqrt 3}{2}\sin\theta_n\cos\varphi_{n_1} -\frac{1}{2}\cos\theta_n\,,
  \\
  \cos\theta_{n_2}
  &=
    +\frac{\sqrt 3}{2}\sin\theta_n\cos\varphi_{n_1} -\frac{1}{2}\cos\theta_n\,,
\end{align}
respectively. The integrands in Eq.~\eqref{eq:angintn} depend on 
$\cos\theta_{n}$, $\cos\theta_{n_1}$, and $\cos\theta_{n_2}$ through the triplet neutron pairing gap $\Delta_n \propto \sqrt{1+3\cos^2\theta }$.

\section{Rotochemical heating with different magnetic field for ordinary pulsars}
\label{sec:magnetic-field}

In Sec.~\ref{sec:results}, we fix the spin-down rate $P\dot P =
10^{-15}\unit{s}$, i.e., the dipole magnetic field $B \sim 10^{12}\unit{G}$, for ordinary pulsars, though it generically takes a value in a rather broad range: $P\dot P = 10^{-17} - 10^{-13}\unit{s}$,
corresponding to $B \sim 10^{11-13}\unit{G}$. Since this value controls the evolution of angular velocity (see Eq.~\eqref{eq:P-Pdot-dipole}), a change in $P\dot{P}$ may affect the intensity of the rotochemical heating. In this appendix, we study the effects of varying $P\dot{P}$ on  the temperature evolution of ordinary pulsars. We fix the other NS parameters to be $M = 1.4\,M_\odot$ and $\Delta M / M = 10^{-15}$ in the following analysis.

\begin{figure*}
  \centering
  \begin{minipage}{0.5\linewidth}
    \includegraphics[width=1.0\linewidth]{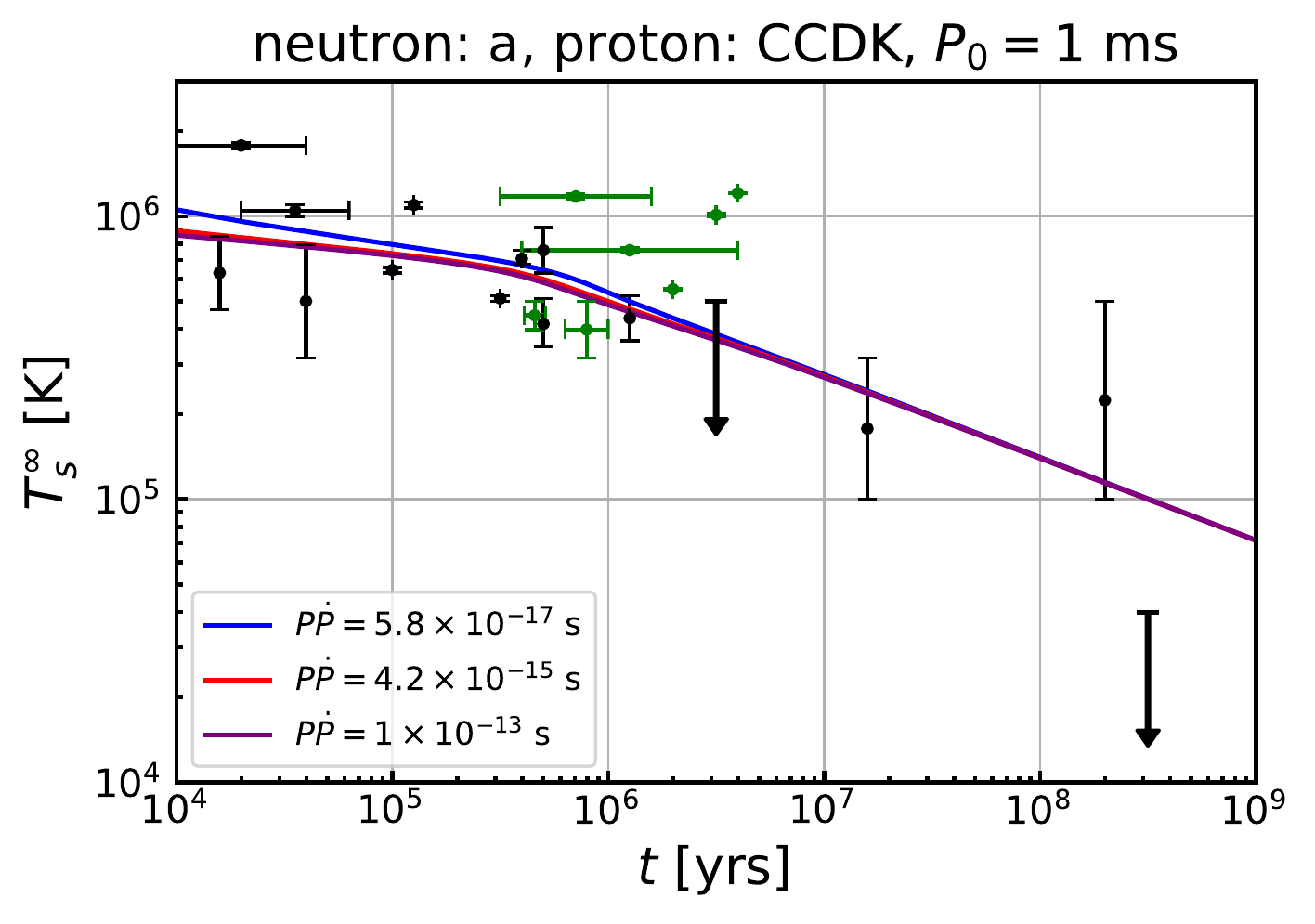}
  \end{minipage}%
  \begin{minipage}{0.5\linewidth}
    \includegraphics[width=1.0\linewidth]{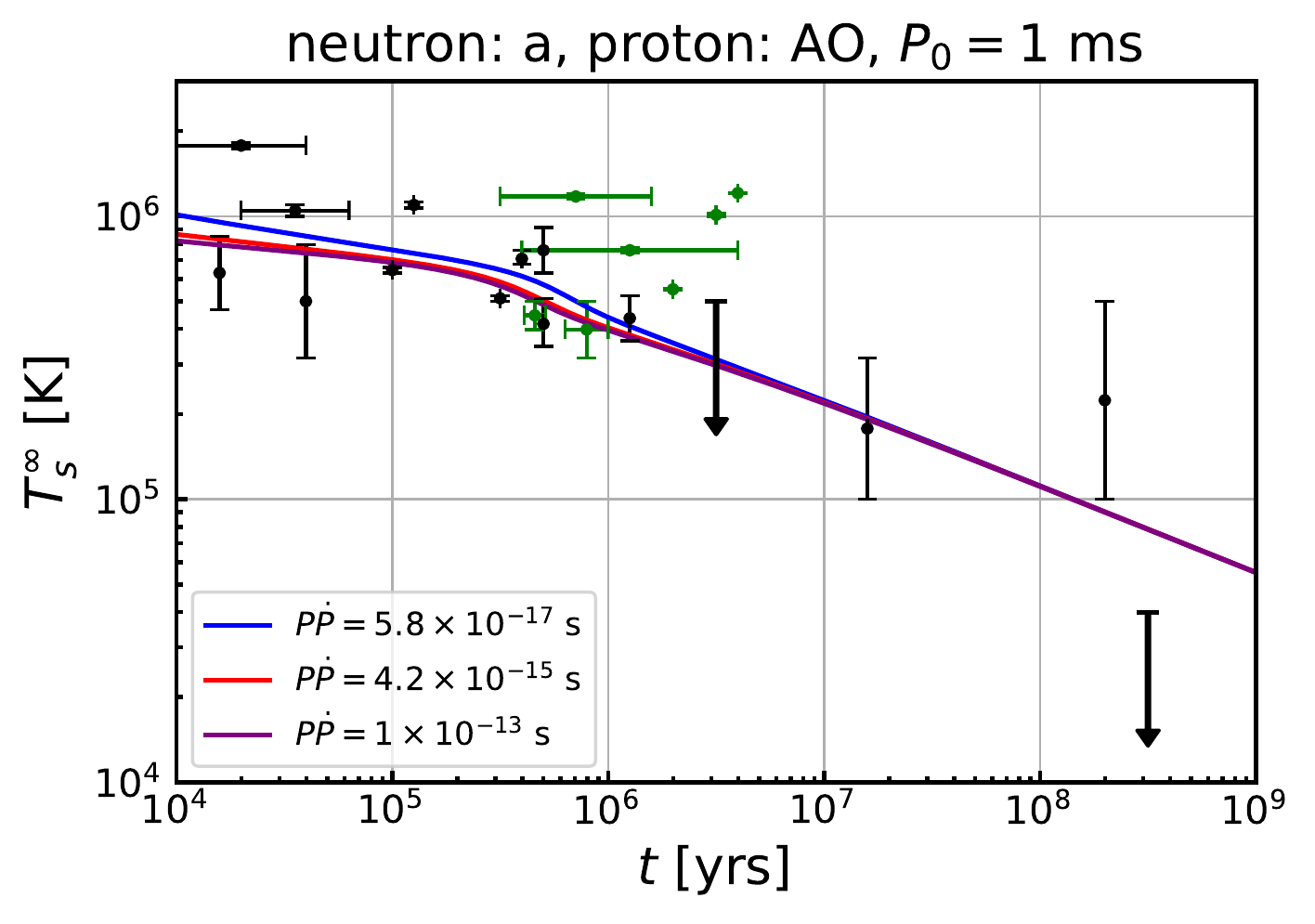}
  \end{minipage}
  \begin{minipage}{0.5\linewidth}
    \includegraphics[width=1.0\linewidth]{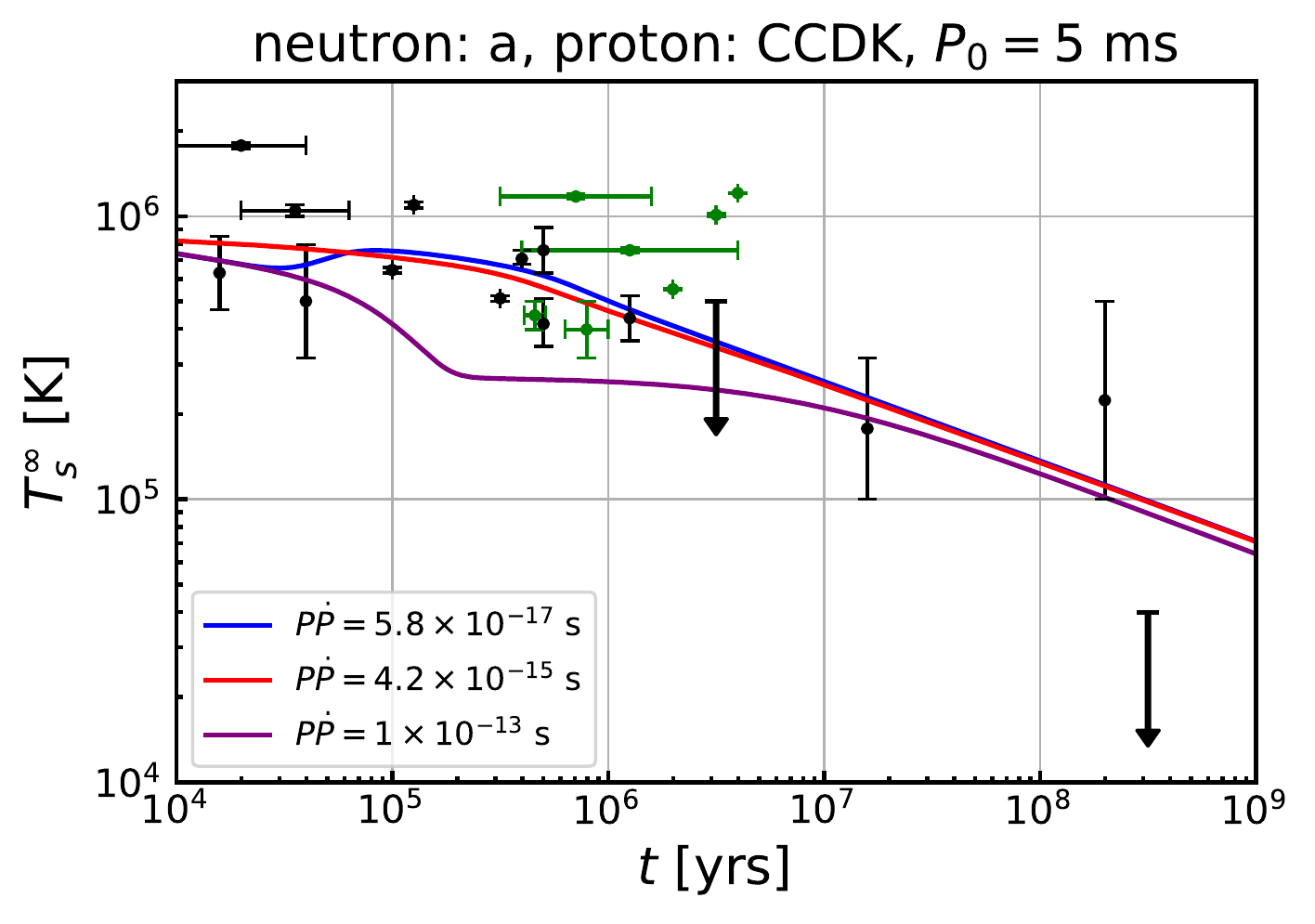}
  \end{minipage}%
 \begin{minipage}{0.5\linewidth}
   \includegraphics[width=1.0\linewidth]{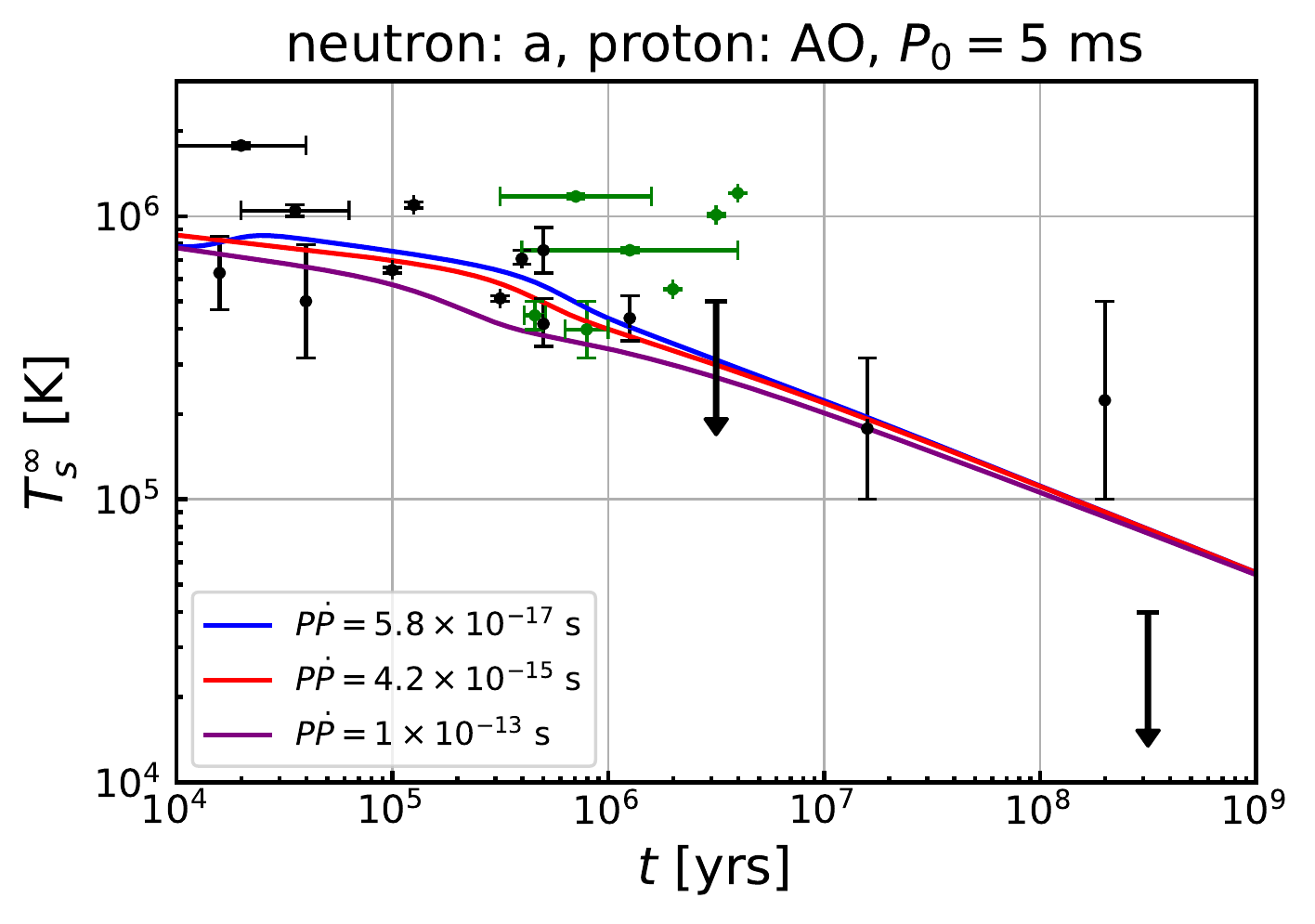}
  \end{minipage}
  \begin{minipage}{0.5\linewidth}
    \includegraphics[width=1.0\linewidth]{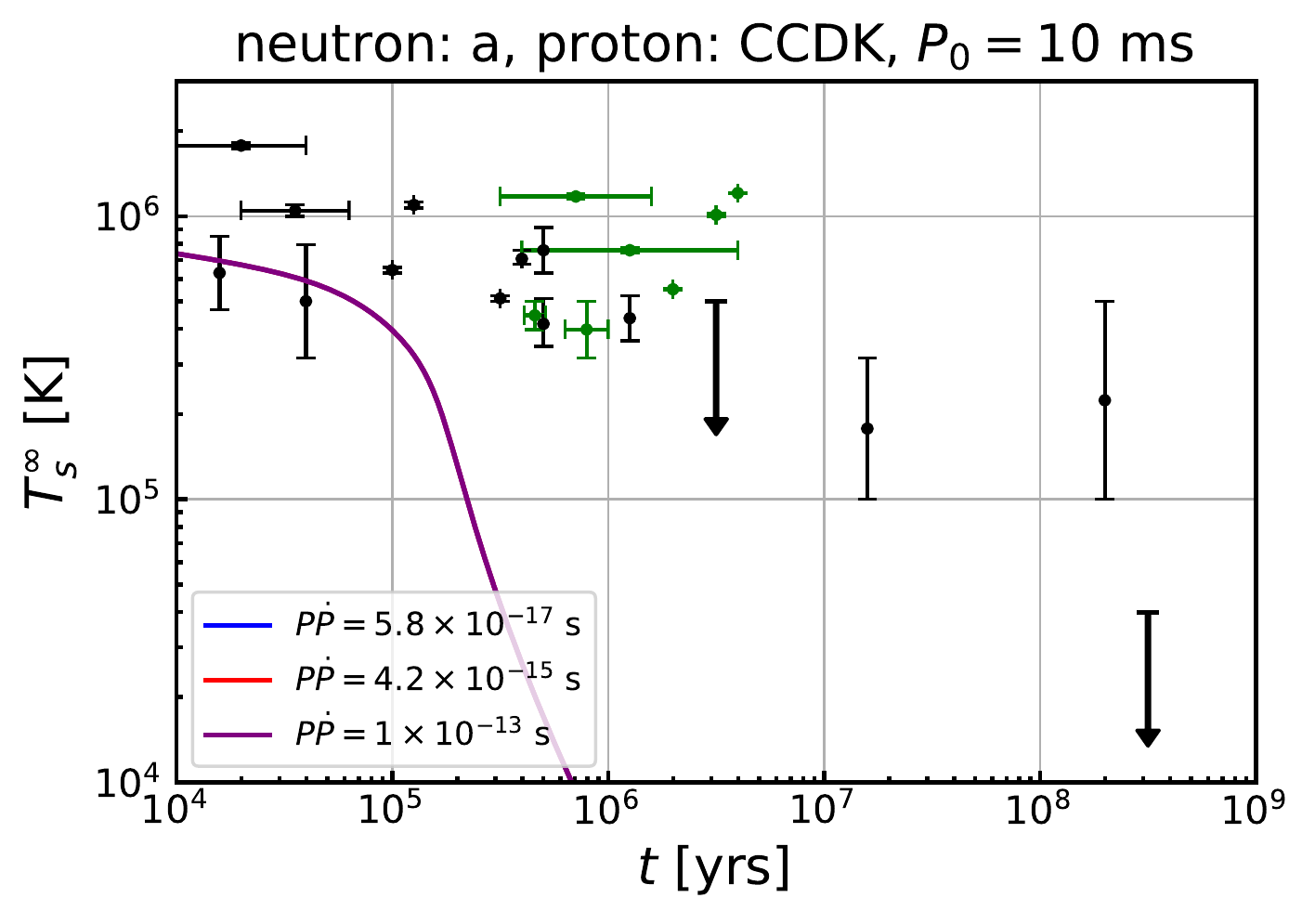}
  \end{minipage}%
  \begin{minipage}{0.5\linewidth}
    \includegraphics[width=1.0\linewidth]{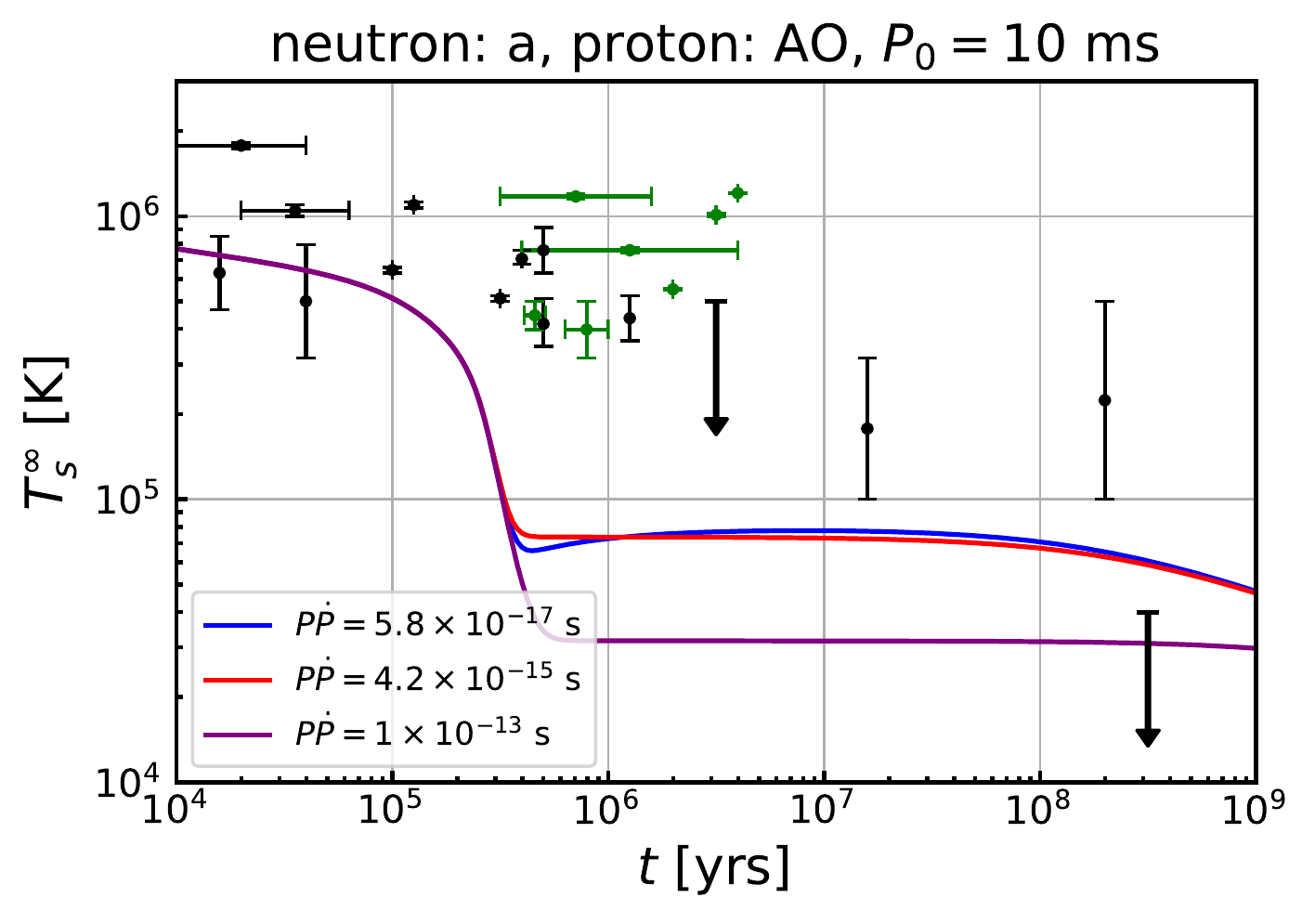}
  \end{minipage}
  \caption{Thermal evolution of ordinary pulsars for different values of $P\dot{P}$.
    We use the neutron ``a'' and proton CCDK (AO) gap models in the left (right) panels.
    The top, middle, and bottom panels show the cases of $P_0=1\unit{ms}$, $5\unit{ms}$, and $10\unit{ms}$, respectively.
    In each panel, we take three values of $P\dot{P}$: $P\dot{P} =5.8\times 10^{-17}\unit{s}$ (blue), $4.2\times10^{-15}\unit{s}$ (red) and $1\times10^{-13}\unit{s}$ (purple).
  The points with error bars are the same as Fig.~\ref{fig:cp-2}, showing the observed surface temperatures.}
  \label{fig:cp-3}
\end{figure*}
Figure~\ref{fig:cp-3} shows the thermal evolution for $P\dot{P} = 5.8\times 10^{-17}\unit{s}$ (blue), $4.2\times10^{-15}\unit{s}$ (red), and $1\times10^{-13}\unit{s}$ (purple), corresponding to $B \sim 10^{11}\unit{G}$, $10^{12}\unit{G}$, and $10^{13}\unit{G}$, respectively. Note that $5.8\times10^{-17}\unit{s}$ is the observed value of $P\dot{P}$ for B0950+08 and $4.2\times10^{-15}\unit{s}$ is that for J2144-3933. We use the neutron ``a'' and proton CCDK (left panels) or AO (right panels) gap models.
The top, middle, and bottom panels show the cases of $P_0 = 1\unit{ms}$,
$5\unit{ms}$, and $10\unit{ms}$, respectively. For $P_0 =1\unit{ms}$
and $5\unit{ms}$, we see that the surface temperature is lower for a larger
$P\dot{P}$ at middle age. This is because with a larger $P\dot{P}$, rotochemical
heating begins earlier, resulting in a smaller $\eta_\ell$ at middle age. For $t \gtrsim 10^8\unit{yr}$, however, the surface temperatures of different $P\dot{P}$ converge to each other since the spin-down effect becomes so small that the evolution of $\eta_\ell$ is
mainly determined by the modified Urca process.
For $P_0 = 10\unit{ms}$, the rotochemical heating never occurs
for the proton CCDK gap model (bottom left). In the case of the proton AO model, on the other hand, the heating effect is small but still visible (bottom right).
Moreover, the difference of surface temperatures at $t\gtrsim 10^6\unit{yr}$ is
larger than that for $P_0 = 1\unit{ms}$ or $P_0 = 5\unit{ms}$.
In this case, the accumulated $\eta_\ell$ is close to the rotochemical
threshold, and hence the heating luminosity is very sensitive to the difference in the $\eta_\ell$, making the temperature evolution highly dependent on other NS parameters such as $P\dot{P}$.

Notice that although a large $P\dot{P}$ (and thus a large dipole magnetic field) tends to
predict low surface temperature, it does not help explain the low
temperature of J2144-3933 compared to that of J0108-1431 and B0950+08, as indicated by the red and blue curves in Fig.~\ref{fig:cp-3}, which are always close to each other.
Consequently, the different temperatures of these NSs should be attributed to the difference in $P_0$.

\bibliographystyle{utphysmod}
\bibliography{rotochemical} 

\end{document}